\documentclass[10pt,aps,prd,showpacs,nofootinbib,floats,floatfix,preprintnumbers,superscriptaddress,twocolumn,showkeys]{revtex4-2}

\usepackage{aas_macros}
\usepackage{cancel}

\usepackage[normalem]{ulem} 
\usepackage{bm}
\usepackage[titletoc]{appendix}
\usepackage{hyperref}
\usepackage{latexsym}
\usepackage{dcolumn}
\usepackage{amsmath,amsfonts,amssymb}
\usepackage{graphicx,epsfig}
\usepackage{amsmath}
\usepackage{fancyhdr}
\usepackage{hyperref}
\usepackage{graphicx,epstopdf}
\usepackage{xcolor}
\usepackage{siunitx}

\def\bea{\begin{eqnarray}}
	\def\eea{\end{eqnarray}}

\begin{document}

\title{Probing horizonless compact objects through observable signatures and images of accretion disk}

\author{Gargi Sen}
\email{sgargip5@gmail.com (Corresponding author)}
\affiliation{Department of Physics, Indian Institute of Technology Guwahati, Assam-781039, India}

\author{Rajibul Shaikh}
\email{rajibulsk@isical.ac.in}
\affiliation{Physics and Applied Mathematics Unit, Indian Statistical Institute, Kolkata-700108, India}

\author{Sayan Chakrabarti}
\email{sayan.chakrabarti@iitg.ac.in}
\affiliation{Department of Physics, Indian Institute of Technology Guwahati, Assam-781039, India}

\author{Santabrata Das}
\email{sbdas@iitg.ac.in}
\affiliation{Department of Physics, Indian Institute of Technology Guwahati, Assam-781039, India}

\date{\today}
	
\begin{abstract}
The observed images of  M87$^*$ and Sgr A$^*$ by Event Horizon Telescope motivate us to study the spacetime geometry
and accretion flow dynamics around compact objects beyond the Kerr/Schwarzschild scenario. In this work, we investigate the properties of a steady, inviscid, axisymmetric, advective accretion flow around the Zipoy-Voorhees ($\gamma$-metric) spacetime by solving the relativistic hydrodynamic equations. The $\gamma$-metric is a static, axisymmetric generalization of the Schwarzschild geometry and is characterized by a mass ($M_\bullet$) and a deformation parameter ($\gamma$). We examine how the deformation parameter $\gamma$ influences the effective potential, the location and nature of the critical points, and the resulting accretion flow topologies. Then, using relativistic thermal bremsstrahlung emissivity, we compute the spectral energy distribution and bolometric luminosity of the flow. We find that $\gamma$ affects the luminosity in opposite ways for the flow having outer critical point (O-type) and that having inner critical point (I-type). Finally, combining the flow properties with general relativistic ray tracing, we construct synthetic images of the accretion disk for different values of $\gamma$ and observer inclination angles. The resulting observed flux distribution independently confirms the trends obtained from the luminosity and spectral analyses. Although the shadow of the $\gamma$-metric  has been shown to be consistent with the observed shadow radii of M87$^*$ and Sgr A$^*$, its accretion flow topology, the resulting radiative properties and image differ from the Schwarzschild case,  thereby giving a complementary way to probe the deviations from the Schwarzschild paradigm.

\end{abstract}

\maketitle
	
\section{Introduction}
The direct imaging of the supermassive black holes M87$^*$ and Sagittarius A$^*$ (Sgr A$^*$) by the Event Horizon Telescope (EHT) Collaboration has provided a powerful probe of gravity in the vicinity of compact objects \cite{EHT-2019I,EHT-2022I}. The morphology of the observed shadow, including its size, shape, and asymmetry encodes information not only about the mass and spin of the compact object but also about the geometry of the surrounding spacetime in the strong-field regime.
In General Relativity (GR), the spacetime around an astrophysical compact object is commonly described by either the static Schwarzschild solution or the stationary Kerr solution. Although the observed images are consistent with the Kerr black hole,  they do not uniquely determine the underlying spacetime geometry or conclusively establish the nature of the central compact object. This motivates to  investigate extensively the  nature of alternative compact object spacetimes including regular black holes, wormholes, naked singularities, and a variety of parametrized deformations.

Among the various exact solutions of Einstein's field equations, the Zipoy-Voorhees spacetime, also known as the $\gamma$-metric, is a simple static and axisymmetric vacuum solution. This is a generalization of the Schwarzschild spacetime. The spacetime is characterized by a mass and a dimensionless deformation parameter $\gamma$. For $\gamma=1$, we get back the Schwarzschild solution and $\gamma \neq 1$ describes a
deformed compact object with prolate or oblate geometry depending on the value of $\gamma$. Being a simple exact vacuum solution, the $\gamma$-metric can offer an interesting framework for studying how deviations from spherical symmetry (Schwarzschild spacetime) affect the motion of accreting matter and the related observable signatures. Most previous works on the $\gamma$-metric have focused on the motion of test particles including timelike and null geodesics ~\cite{Chowdhury-etal2011, Boshkayev-etal2015, Toshmatov-Malafarina2019},
quasinormal modes~\cite{Allahyari-etal2018}, shadows~\cite{Shaikh-etal2021}, gravitational lensing~\cite{Chakrabarty-etal2021}. While these studies tell us a lot about the shape and structure of the spacetime, they do not account the exact mechanisms to produce electromagnetic radiation observed by a distant observer.  In realistic astrophysical systems, compact objects are embedded in hot accreting plasma, and therefore,
the observed radiation coming from the vicinity of a compact object depends not only on the spacetime
geometry but also on the dynamics and thermodynamic
properties of the accretion flow. Moreover, the horizon-scale images of M87$^*$ and Sgr A$^*$ suggest that these sources are accreting at sub-Eddington rates, which can be well described by radiatively inefficient hot accretion flows characterized by hot, geometrically thick, and optically thin plasma~\cite{EHT-2019V, EHT-2019VI,EHT-2022V,Chatterjee-etal2022}. Relativistic hydrodynamic accretion provides a physically self-consistent framework for connecting the background spacetime with observable quantities. The spacetime geometry determines the effective potential experienced by the flow, which in turn governs the accretion flow dynamics. The dynamical features strongly influence the density and temperature distribution of the
accreting plasma and consequently modify the emitted
radiation. Therefore, a realistic model of the accretion flow is essential for properly accounting for the emission processes occurring within the flow. Thermal bremsstrahlung emission constitutes one of the dominant radiative mechanisms in optically thin, geometrically thick hot accretion flows \cite{Narayan-Yi1995b,Mahadevan1998}. Since the bremsstrahlung emissivity depends on the density and temperature of the flow, it naturally carries information about both the accretion dynamics and the underlying spacetime geometry. By incorporating such physically motivated emissivity combined with general relativistic ray tracing, we can construct more realistic synthetic images than those obtained from idealized emissivity prescriptions or purely geodesic analyses.

The theoretical prediction of a black hole shadow against the surrounding emission was proposed well before its recent observational realization \cite{Falcke-etal1999}. Using general relativistic ray tracing, the synthetic images of Sgr A$^*$ were simulated to investigate the appearance and apparent size of the shadow produced by photon capture and gravitational light bending. This theoretical picture gained observational support nearly two decades later, when EHT obtained horizon-scale images of M87$^*$ and Sgr A$^*$~\cite{EHT-2019I,EHT-2022I}.
During this period, significant progress was made in modeling the dynamics of magnetized accretion flows through general relativistic magnetohydrodynamic (GRMHD) simulations to develop a numerical scheme that is capable of self-consistently evolving turbulent, magnetized plasma in strong gravitational fields~\cite{Gammie-etal2003}. After that, combining the GRMHD simulation with general relativistic radiative transfer, the
dynamically self-consistent synthetic images and spectra of the emitting plasma were generated~\cite{Noble-etal2007, Moscibrodzka-etal2009}. This approach was further refined via the development of general relativistic radiative transfer codes, such as GRTRANS~\cite{Dexter2016}, IPOLE~\cite{Moscibrodzka-Gammie2018}, BHOSS~\cite{Younsi-etal2012, Younsi-etal2016} and was used to model the horizon-scale emission from M87$^*$ and Sgr A$^*$~\cite{Shcherbakov-etal2012, Moscibrodzka-etal2015, Younsi-etal2019}. These simulations self-consistently capture the time dependent evolution of the magnetic field, plasma density, temperature, and velocity fields, which is combined with radiative transfer to predict the observed image. However, GRMHD simulations are computationally expensive and are designed to capture how accretion flows evolve over time. In contrast, a steady state hydrodynamic treatment provides an alternative approach in which the global structure of the accretion flow can be obtained directly from the relativistic conservation equations. This framework allows us to systematically investigate the connection among the background spacetime geometry, the structure and flow properties of the accreting plasma, and the resulting observational signatures. This motivates to probe the possible deviations from the Kerr/Schwarzschild black holes using horizon-scale imaging features. One widely used method that avoids the high computational cost of full GRMHD simulations is a simplified steady state model built on the RIAF framework~\cite{Narayan-Yi1994,Yuan-etal2003,Broderick-etal2011, Younsi-etal2012}, where the density and temperature follow a prescribed profile rather than being obtained by solving the underlying equations of motion. Therefore, our motivation for this work is to build a steady state accretion flow around the $\gamma$-metric by directly solving the hydrodynamic equations, and use it to generate synthetic images that allow us to distinguish this deformed spacetime from the Schwarzschild case. Within our model framework, we obtain global transonic accretion solutions around this deformed spacetime for a given set of flow parameters (specific energy and specific angular momentum) and parameters of the central object. We consider a relativistic equation of state to accurately describe the thermodynamic properties of the accreting matter. The radiative properties such as spectra and luminosity are modeled with a relativistically corrected thermal bremsstrahlung emissivity. After that, we investigate how the accretion flow varies with $\gamma$ and the different accretion solutions it produces. 
 We then combine this flow structure with general relativistic ray tracing technique to generate synthetic horizon-scale images. This shows how the $\gamma$-metric differs from the Schwarzschild one, and how spacetime deformation affects the observable properties of the accretion flow. It has been shown that the parameter range $\gamma\geq1/2$ for which the $\gamma$-metric casts shadows and hence mimics a black hole is consistent with the observed shadow radii of both M87$^*$~\cite{Shaikh-etal2021} and Sgr A$^*$~\cite{Shaikh-2022}. One of our aims in this paper is to study whether observational aspects related to the accretion flow can distinguish the $\gamma$-metric from a Schwarzschild black hole.
 
The paper is organized as follows. In Section~\ref{sec:Geometry}, we describe the background geometry of the $\gamma$-metric. In Section~\ref{sec:gov_equations}, we present the model assumptions and the governing hydrodynamic equations for a steady, inviscid, advective accretion flow. In Section~\ref{sec:CP_analysis}, we discuss the procedure of identifying the location of critical point, analyze the nature of them as a function of the flow parameters and the deformation parameter ($\gamma$). Then we obtain different types of transonic solutions in this spacetime. In Section~\ref{sec:parameter_space}, we examine the parameter space corresponding to the different accretion flow topologies. In Section~\ref{sec:Rad_properties}, we study the radiative properties of the flow, including the spectral energy distribution, the bolometric luminosity, and the synthetic images of the accretion disk as observed by a distant observer. Finally, we summarize our main findings with limitations in Section~\ref{sec:Conclusions}.

\section{Background Geometry}
\label{sec:Geometry}
In this work, we consider the Zipoy-Voorhees spacetime
\cite{Zipoy1966,Voorhees1970}, commonly known as the
$\gamma$-metric, which belongs to the Weyl class of static,
axisymmetric vacuum solutions of Einstein's field equations.
The $\gamma$-metric is one of the simplest exact generalizations of the Schwarzschild spacetime and provides a useful framework for investigating how deviations from the Schwarzschild black hole influence accretion flows and their observable properties. In Erez-Rosen coordinates,
the spacetime line element is given by,

\begin{align}
\label{eq:line}
ds^2 &= -A(r)\,dt^2
+\frac{1}{A(r)}
\Big[B(r,\theta)\,dr^2
+C(r,\theta)\,d\theta^2
\notag\\
&\qquad
+\left(r^2-2mr\right)\sin^2\theta\,d\phi^2
\Big],
\end{align}

where the metric functions are

\begin{align}
A(r) &=
\left(1-\frac{2m}{r}\right)^{\gamma},
\notag\\
B(r,\theta)
&=
\left[
\frac{r^2-2mr}
{r^2-2mr+m^2\sin^2\theta}
\right]^{\gamma^2-1},
\notag\\
C(r,\theta)
&=
\frac{(r^2-2mr)^{\gamma^2}}
{\left(r^2-2mr+m^2\sin^2\theta\right)^{\gamma^2-1}}.
\label{eq:metric_function}
\end{align}

The spacetime is characterized by two independent parameters,
namely the mass parameter $m$ and the dimensionless deformation
parameter $\gamma$. The parameter $m$ is related to the mass of
the compact object, whereas $\gamma$ quantifies the deviation of
the spacetime from spherical symmetry. The corresponding
Arnowitt--Deser--Misner (ADM) mass is given by
$M_\bullet=m\gamma$, while the mass quadrupole moment is $Q=\frac{\gamma M_\bullet^3(1-\gamma^2)}{3}$, for $\gamma\neq1$ \cite{Herrera-etal1998}. The deformation parameter ($\gamma$) governs the geometry of the
spacetime. In the limiting case $\gamma=1$, the metric reduces
to the Schwarzschild solution, whereas $\gamma=0$ corresponds to
the flat Minkowski spacetime. For all other values of $\gamma$,
the spacetime remains static and axially symmetric but departs
from spherical symmetry. In particular, the geometry becomes
prolate for $\gamma<1$ and oblate for $\gamma>1$. Since the $\gamma$-metric is an exact vacuum solution of
Einstein's field equations, both the Ricci tensor ($R_{\mu\nu}$), and the Ricci scalar ($R=g^{\mu\nu}R_{\mu\nu}$) vanish identically. However, the
Kretschmann scalar $(K=R_{\mu\nu\alpha\beta}R^{\mu\nu\alpha\beta})$ reveals the existence of curvature singularities. One
singularity is located at $r=0$, for $\gamma>0$. A second curvature singularity occurs at $r_{\rm s}=2m=\frac{2M_\bullet}{\gamma}$, where the Kretschmann scalar diverges for all values of
$\theta$ whenever $0<\gamma<2$, except for the special cases
$\gamma=0$ and $\gamma=1$. Therefore, the hypersurface
$r_{\rm s}=2m$ represents a curvature singularity rather than coordinate singularity. The Schwarzschild spacetime
($\gamma=1$) is exceptional, for which the surface $r_{\rm s}=2m$
corresponds to the event horizon and this singularity
is merely a coordinate singularity or removable singularity. Furthermore, the surface
$r_{\rm s}=2m$ is infinitely redshifted with respect to an observer
located at spatial infinity. In our present work, we restrict our analysis to $\gamma\geq0.5$ for the reason mentioned in the Introduction.

Throughout this work, we concentrate our analysis to the region
outside the outermost singular surface, i.e.,
$r>2m$. Consequently, the curvature singularities enclosed
within this radius do not directly influence the accretion flow
considered here. We adopt the metric signature $(-,+,+,+)$ and employ
geometrized units by setting $G=c=M_\bullet=1$, where $G$ is the gravitational constant and $c$ is the speed of light in vacuum. In this system of units,
distances and specific energies are measured in units of
$r_{\rm g}=GM_\bullet/c^2$ and $c^2$, respectively. Since we focus on
geometrically thin accretion flows confined to the disk plane,
all calculations are performed on the equatorial plane by
setting $\theta=\pi/2$.

\section{Model Assumptions and governing equations}
\label{sec:gov_equations}

We consider a low angular momentum, steady, inviscid, axisymmetric, advective accretion flow around a  Zipoy-Voorhees spacetime. We assume that the accretion disk lies on the equatorial plane of the central object, and in the absence of viscosity, radiative cooling, and magnetic fields, the flow conserves both energy and angular momentum. 

Therefore, the conservation of energy-momentum and mass flux equations for the perfect fluid in hydrodynamic framework are given by,
\begin{equation}
    T^{\mu\nu}_{;\nu} = 0 \quad\text{and} \quad (\rho u^{\nu})_{;\nu} = 0,
    \label{coneqn}
\end{equation}
where $T^{\mu\nu} = (e + p)u^{\mu}u^{\nu} + pg^{\mu\nu}$ is the energy-momentum tensor and $j^{\mu} = \rho u^{\mu}$ is the four current of the fluid. 
Here, $e$, $p$, $\rho$, and $u^\mu$ denote the total energy density including rest mass energy, pressure, mass density and four velocities of the fluid, respectively and the spacetime indices $\mu$ and $\nu$ run from $0$ to $3$.
	
The time-like four velocity satisfies the normalization condition $u_\mu u^\mu =-1$. Now, we define the projection operator
$h^{i}_{\mu}=\delta^i_{\mu}+u^iu^{\mu}$, which projects any tensor onto the three-dimensional hypersurface orthogonal to the fluid four velocity by satisfying the condition $h^{i}_{\mu} u^\mu =0$, where $i$ stands for $r,\theta,\phi$.
Therefore, by projecting the energy-momentum conservation equation onto this hypersurface, we obtain the relativistic Euler equation for the fluid motion as,
\begin{equation}
    h^{i}_{\mu} T^{\mu\nu}_{;\nu} = (e+p)u^{\nu}u^{i}_{;\nu} + (g^{i \nu}+u^{i}u^{\nu})p_{,\nu}=0.
    \label{euler}
\end{equation}

Additionally, we project the energy-momentum conservation equation along $u^\mu$ and obtain the entropy generation equation as, 
\begin{equation}
    u_{\mu} T^{\mu\nu}_{;\nu}= u^{\nu}\left[\left(\frac{e + p}{\rho}\right)\rho_{,\nu}-e_{,\nu}\right]=0.
    \label{entgen}
 \end{equation}

The azimuthal velocity is defined as $v^{2}_{\phi}=u^{\phi}u_{\phi}/-u^{t}u_{t}$ and the corresponding Lorentz factor is $\gamma^{2}_{\phi}=1/(1-v^{2}_{\phi})$. The polar velocity is $v_{\theta}^2=\gamma^{2}_{\phi}(u^{\theta}u_{\theta})/(-u^{t}u_{t})$ and polar Lorentz factor is $\gamma^{2}_{\theta}=1/(1-v_{\theta}^2)$. As we consider the flow to be confined on equatorial plane and hence, $\theta=\pi/2$ which leads to maintain $u^\theta=0$ and $\gamma_{\theta}=1$.
Therefore, the three radial velocity of the fluid in the co-rotating frame ~\cite{Lu-1985} is defined as $v^{2}=\gamma^{2}_{\phi}v^{2}_{r}$, where $v^{2}_{r}=(u^{r}u_{r})/(-u^{t}u_{t})$. The radial Lorentz factor is $\gamma^{2}_{v} = 1/(1-v^{2})$.  The symmetric nature of the background spacetime allows to model the accretion flow as stationary and axisymmetric. There exist two commuting Killing
vectors associated with time ($\eta^t$) and azimuthal coordinate
($\eta^\phi$). Therefore, using Killing condition $\nabla_{\mu} \eta_{\nu}+\nabla_{\nu} \eta_{\mu}=0$, we get the specific energy $\mathcal{E}$ and the conserved
angular momentum $\mathcal{L}$ associated with the two Killing vectors $\eta^t$ and $\eta^\phi$, respectively and are given by,

\begin{equation}
    -h u_{t} = {\cal E}  \quad; \quad h u_{\phi} = {\cal L},
    \label{energy-angularmomentum}
\end{equation}

where, $\mathcal E$ is the Bernoulli constant ($i.e.,$ specific energy) and $\cal L$ is the bulk specific angular momentum of the flow. As, we are dealing with the inviscid flow, we can define the specific angular momentum of the flow as $\lambda = {\cal L}/{\cal E}=-u_{\phi}/u_{t}$, which is also a conserved quantity.

 Now, using the definitions of velocities, we get the radial component of the momentum equation from Eq.~\ref{euler} for $i = r$ as,
\begin{equation}
    v\gamma_{v}^{2}\frac{dv}{dr}+\frac{1}{h\rho}\frac{dp}{dr}+\frac{d\Phi^{\textrm{eff}}_{\rm e}}{dr}=0,
    \label{eulerradial}
\end{equation}
where $h\left[ =(e+p)/\rho\right]$ is the specific enthalpy for the flow, and $\Phi^{\rm eff}_{\rm e}$ denotes the effective potential \cite{Dihingia-etal2018} at the disk equatorial plane, which is given by,

\begin{equation}
    \Phi^{\textrm{eff}}_{\rm e}=1+\frac{1}{2}\ln\left[\frac{(r^2-2 m r)A(r)}{r^2-2 m r-\lambda^2A(r)^2}\right].
    \label{potential}
\end{equation}

\begin{figure}[h!]
    \begin{center}
    \includegraphics[width=\columnwidth]{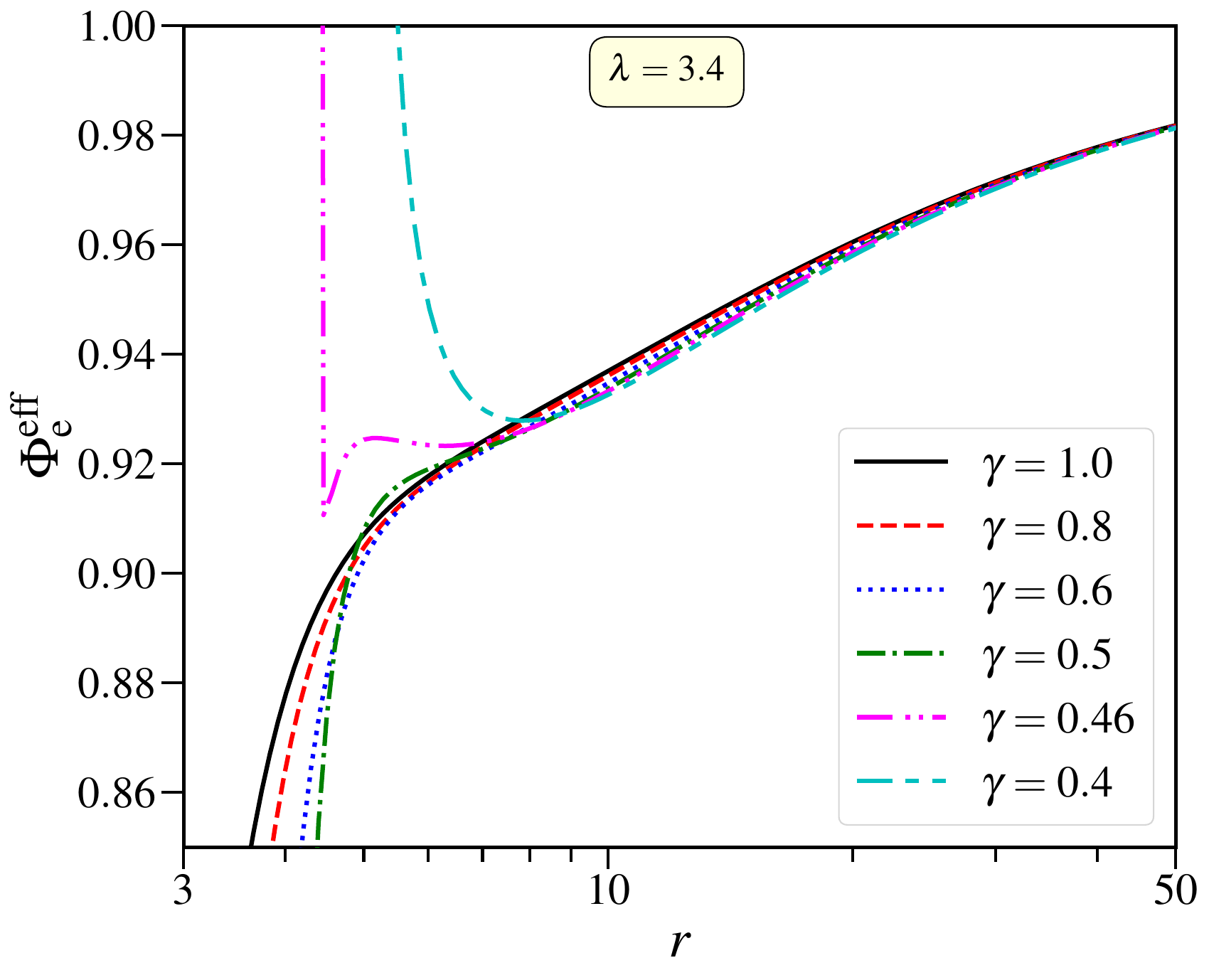}
    \end{center}
    \caption{Radial dependence of effective potential ($\Phi^{\rm eff}_{\rm e}$) for a fixed angular momentum $\lambda=3.4$. Here, solid (black), dashed (red), dotted (blue), dot-dashed (green), dot-dot-dashed (magenta), and small-big-dashed (cyan) curves correspond to the deformation parameter $\gamma=$ 1.0, 0.8, 0.6, 0.5, 0.46, and 0.4, respectively. See the text for details.
    }
    \label{fig:fig_poten1}
\end{figure}

\begin{figure}[h!]
    \begin{center}
    \includegraphics[width=\columnwidth]{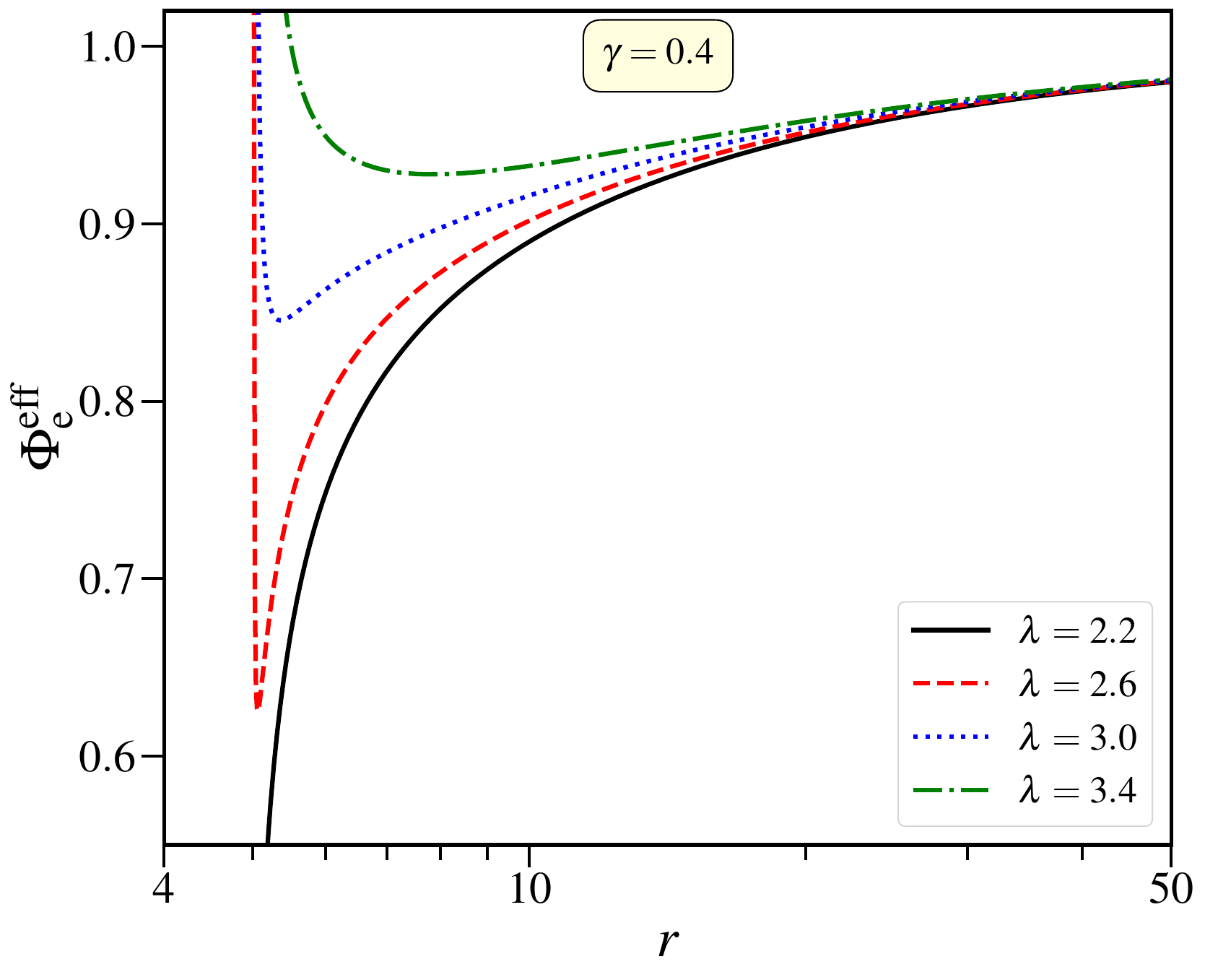}
    \end{center}
    \caption{Effective potential ($\Phi^{\rm eff}_{\rm e}$) as a function of radial distance for $\gamma=0.4$. The solid (black), dashed (red), dotted (blue), and dot-dashed (green) curves represent results for $\lambda=2.2$, 2.6, 3.0, and 3.4, respectively. See the text for details.
    }
    \label{fig:fig_poten2}
\end{figure}

\begin{figure}[h!]
    \begin{center}
    \includegraphics[width=\columnwidth]{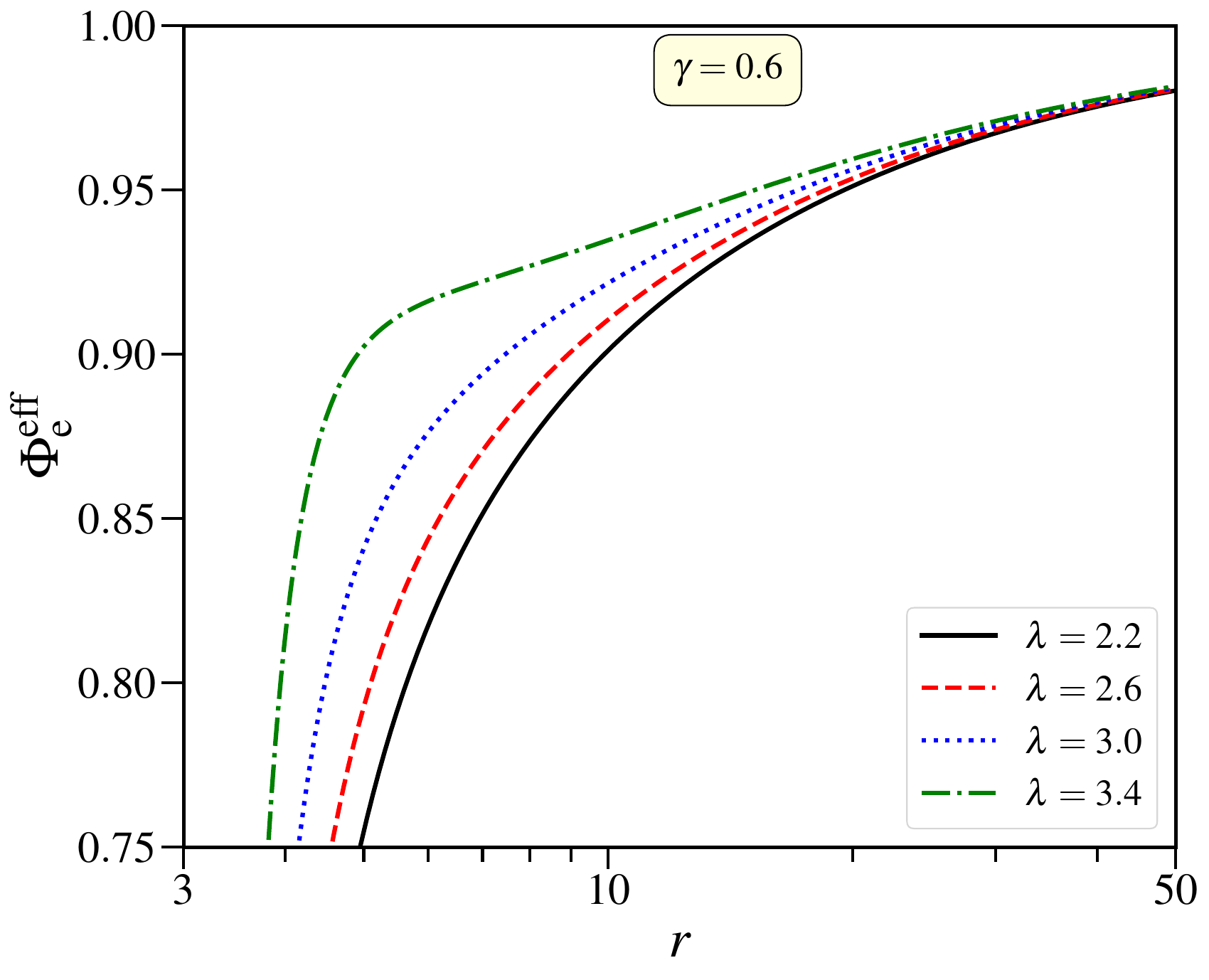}
    \end{center}
    \caption{Variation of effective potential ($\Phi^{\rm eff}_{\rm e}$) with radial distance for $\gamma=0.6$. Here, solid (black), dashed (red), dotted (blue) and dot-dashed (green) curves denote results obtained for $\lambda=2.2, 2.6,3.0$ and $3.4$, respectively. See the text for details.
    }
    \label{fig:fig_poten3}
\end{figure}

The effective potential ($\Phi^{\textrm{eff}}_{\rm e}$) is solely dependent on the spacetime metric and angular momentum of the flow. In Fig.~\ref{fig:fig_poten1}, we illustrate the variation of the $\Phi^{\textrm{eff}}_{\rm e}$ with different values of $\gamma$ while keeping the specific angular momentum fixed at $\lambda=3.4$. The solid (black), dashed
(red), dotted(blue), dot-dashed (green), dot-dot-dashed (magenta), and small-big-dashed (cyan) curves correspond to
$\gamma=1.0, 0.8, 0.6, 0.5, 0.46,$ and $0.4$, respectively. As the $\gamma$ parameter decreases from $1$ (corresponding to the Schwarzschild spacetime) toward $0$ (the flat spacetime limit), the effective potential gradually approaches the Newtonian form. Fig.~\ref{fig:fig_poten2}, represents the variation of the $\Phi^{\textrm{eff}}_{\rm e}$ for different values of $\lambda$ with $\gamma=0.4$. The solid (black), dashed
(red), dotted (blue) and dot-dashed (green) curves are for $\lambda=2.2, 2.6,3.0,$ and $3.4$, respectively. Next we examine the corresponding effective potential profiles for $\gamma=0.6$ using the same set of angular momentum values in Fig.~\ref{fig:fig_poten3}. 
A comparison between Figs.~\ref{fig:fig_poten2} and \ref{fig:fig_poten3} reveal that while decreasing the $\gamma$ parameter, the centrifugal barrier in the inner region becomes stronger, which causes the effective potential to differ progressively from that of the Schwarzschild black hole.

 We integrate 2nd part of Eq.~\ref{coneqn} to obtain another constant of motion, which is known as the mass accretion rate ($\dot M$) and is given by,
 \begin{equation}
     \dot{M}=4\pi\rho v \gamma_v \sqrt{\frac{C(r,\pi/2)\times(r^2-2 m r)}{A(r)}}\frac{H}{r}. 
     \label{mdot}
 \end{equation}
 
 In our work, we use the dimensionless form of mass accretion rate as ${\dot m}={\dot M}/{\dot M}_{\rm Edd}$, where ${\dot M}_{\rm Edd} ~(=1.44 \times 10^{18}(M_\bullet/M_\odot)$ gm s$^{-1}$) is the Eddington accretion rate and $M_\odot$ is the solar mass. In Eq.~\ref{mdot}, $H$ refers to the local half-thickness of the accretion disk. Assuming the flow is in hydrostatic equilibrium in the vertical direction,  $H$ is given by~\cite{Riffert-Herold1995, Peitz_Appl1997}
 \begin{equation}
     H=\sqrt{\frac{pr^{3}}{\rho F}}; \quad F=\frac{1}{1-\lambda\Omega},
     \label{height}
 \end{equation}
 where $\Omega ~ [=(\lambda A(r)^2)/(r^2 - 2 m r)]$ is the angular velocity of the fluid.

As the temperature of the accretion flow close to the
compact object reaches upto of the order $\sim10^{10-12}$ K, the flow becomes thermally relativistic. Therefore, to close the Eqs.~\ref{coneqn} and \ref{mdot}, we use the relativistic equation
of state (REoS)~\cite{Chattopadhyay-Ryu2009} with variable adiabatic index that relates total local energy density ($e$), pressure ($p$) and mass density ($\rho$) as,
 \begin{equation}
     e = \frac{\rho f}{\tau}, \quad p=\frac{2\rho\Theta}{\tau},
     \label{eos}
 \end{equation}
 with $\tau = 1+ m_p/m_e$ and 
 $$
 f=\left[ 1+ \Theta \left( \frac{9\Theta +3}{3 \Theta +2}\right) \right] + \left[ \frac{m_p}{m_e} + \Theta \left( \frac{9\Theta m_e +3m_p}{3 \Theta m_e+ 2 m_p}\right) \right],
 $$
 where $m_p$ and $m_e$ denote the masses of ion and electron, respectively, and we use dimensionless temperature as $\Theta=k_{\rm B}T/m_e c^2$. According to the REoS, the sound speed can be expressed as $C_s = \sqrt{2\Gamma\Theta/(f+2\Theta)}$, where $\Gamma ~ [= (1+N)/N]$ represents the adiabatic index and $N ~[=(1/2)(df/d\Theta)]$ corresponds to the polytropic index of the flow.  By integrating Eq.~\ref{entgen}, we can calculate the density of the flow as, 
 
 \begin{equation}
     \rho=\mathcal{K}\exp(k_1) \Theta^{3/2} \left(2 + 3 \Theta\right)^{3/4} \left(3 \Theta + \frac{2m_p}{m_e}\right)^{3/4},
     \label{density}
 \end{equation}
 where $\mathcal{K}$ is the constant of entropy and 
 $k_1 = \left[ f - \left(1+m_p/m_e\right)\right]/2\Theta$. Now, using Eqs.~\ref{mdot}
and \ref{density}, we obtain the measure of entropy by calculating the entropy accretion rate ($\dot{\mathcal{M}}$) \cite{Kumar-Chattopadhyay2017}, which is given by,
 \begin{align}
\dot{\mathcal{M}}=&\frac{\dot{M}}{4 \pi \mathcal{K}}=\exp(k_1) \Theta^{3/2} \left(2 + 3 \Theta\right)^{3/4} \left(3 \Theta + \frac{2m_p}{m_e}\right)^{3/4}\nonumber\\
&\times v \gamma_v \sqrt{\frac{C(r,\pi/2)\times(r^2-2 m r)}{A(r)}}\frac{H}{r}
\label{entacc}
 \end{align}
    
 Note that for a non-dissipative flow characterized with a given set of energy ($\mathcal{E}$) and angular momentum ($\lambda$), $\dot{M}$ and $\dot{\mathcal{M}}$ remain conserved all throughout from outer edge ($r_{\rm edge}^{\rm out}$) to inner edge ($r_{\rm edge}^{\rm in}$) of the disk.

 Therefore using Eqs.~\ref{eulerradial}, \ref{entgen}, \ref{energy-angularmomentum}, \ref{mdot} and \ref{eos}, we obtain the wind equation as,
\begin{equation}
    \frac{dv}{dr}=\frac{\mathcal{N}}{\mathcal{D}},
    \label{dvdr}
\end{equation}
where the numerator ($\mathcal{N}$) is given by,

\begin{equation}
\mathcal{N}=\frac{2 C^{2}_{s}}{\Gamma +1}
\left(\mathcal{N}_1\right)-\frac{d\Phi^{\rm eff}_{\rm e}}{dr},
\label{numerator}
\end{equation}

\begin{equation}
    \mathcal{N}_1=\frac{1}{2r}-\frac{1}{2F}\frac{dF}{dr}
+\frac{r^2+m^2\gamma(1+\gamma)-m r (2+\gamma)+(m-r)^2}
{(m-r)(2m-r)r}
\end{equation} and the denominator ($\mathcal{D}$) is given by, 
\begin{equation}
    \mathcal{D}=\gamma_{v}^{2}\left(v-\frac{2 C^{2}_{s}}{v(\Gamma +1)}\right).
    \label{denominator}
\end{equation}
Further, using Eqs.~\ref{mdot}, \ref{height} and \ref{dvdr}, we calculate the radial gradient of the dimensionless temperature as,
\begin{equation}
    \frac{d\Theta}{dr}=\frac{-2\Theta}{2N+1}\left(\mathcal{N}_1\right).
    \label{dthetadr}
\end{equation}

\section{Critical points analysis and transonic solutions}
\label{sec:CP_analysis}
In the accretion process, the rotating matter starts its journey from $r_{\rm edge}^{\rm out}$ with a subsonic speed $(v<C_s)$ and progressively accelerates due to the strong gravitational attraction of the compact object as the matter approaches toward them. The accretion flow around $\gamma$- metric remains smooth everywhere, and consequently the radial velocity gradient of the flow $(dv/dr)$ must be real and finite throughout the flow. Nevertheless, the Eq.~\ref{denominator} clearly suggests that the denominator ($\mathcal{D}$) may vanish at some points. To maintain regularity of the flow, the numerator ($\mathcal{N}$) must also vanish simultaneously at those points. Such specific points, where $\mathcal{N}=\mathcal{D}=0$, are called as critical points ($r_{\rm c}$). Therefore, imposing the condition $\mathcal{D}=0$, we obtain the radial velocity ($v_{\rm c}$) of the flow at the critical point ($r_{\rm c}$) as,

\begin{equation}
    v_{\rm c}=\sqrt{\frac{2 }{\Gamma_{\rm c} +1} }C_{\rm sc}.
    \label{vc}
\end{equation}
Similarly, the condition $\mathcal{N}=0$ yields the sound speed ($C_{\rm sc}$) at $r_{\rm c}$ as,
\begin{equation}
    C^{2}_{\rm sc}=\frac{\Gamma_{\rm c} +1}{2}\left(\frac{d\Phi^{\textrm{eff}}_{\rm e}}{dr}\right)_{\rm c}
    \left(\mathcal{N}_1\right)_{\rm c}^{-1}.
    \label{csc}
\end{equation}
In Eqs.~\ref{vc} and \ref{csc}, quantities with subscript `c' denote the values at the critical point ($r_{\rm c}$). From Eq.~\ref{csc} we can calculate the temperature of the flow at $r_{\rm c}$.

\begin{figure}[h!]
    \begin{center}
    \includegraphics[width=\columnwidth]{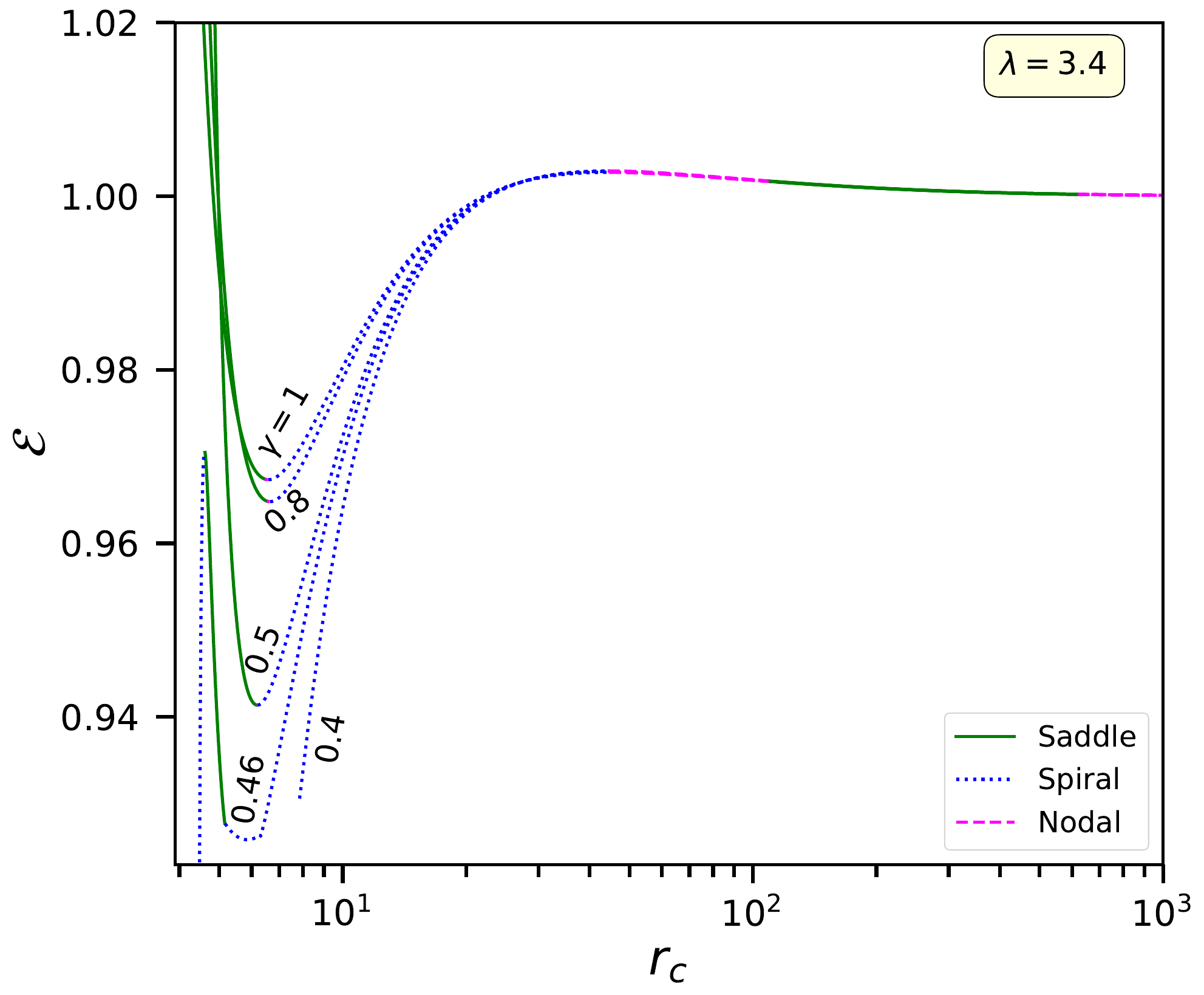}
    \end{center}
    \caption{Variation of energy ($\mathcal{E}$) with critical point location ($r_c$) for $\lambda=3.4$. The solid (green), dotted (blue), and dashed (magenta) segments in a given curve correspond to saddle, spiral, and nodal types critical points, respectively. Different deformation parameters ($\gamma=1.0$, 0.8, 0.5, 0.46, and 0.4) are marked in the figure. See the text for details.}
    \label{fig:energy1}
\end{figure}

As the flow is smooth throughout the streamline, including at the critical points ($r_{\rm c}$), the radial velocity gradient $(dv/dr)$ is required to be real and finite at all radial points. Therefore, we employ L’Hôpital’s rule at the critical point to avoid indeterminate form of the radial velocity gradient $(dv/dr=0/0)$ at $r_{\rm c}$, and evaluate the value of $(dv/dr){\rm c}$ at the critical point. For a given set of input parameters ($\mathcal E$, $\lambda$, and $\gamma$), $(dv/dr)_{\rm c}$ admits two values. When both $(dv/dr)_{\rm c}$ are real and possess opposite signs, the corresponding critical point is classified as a saddle type. In contrast, when $(dv/dr)_{\rm c}$ are real and have the same sign, the critical point is of nodal type, while for spiral type critical points $(dv/dr)_{\rm c}$ are imaginary. Moreover, the saddle type critical points are stable against perturbation \cite{Kato-etal1993} in contrast to the unstable nodal and spiral types critical points, they are of particular astrophysical importance, as transonic accretion solutions can pass only through them. Hereafter, unless stated otherwise, we refer exclusively to saddle-type critical points as critical points. The sign of $(dv/dr)_{\rm c}$ determines the nature of the flow, where negative and positive values correspond to accretion and wind solutions, respectively. Here, we consider only the accretion branch. Additionally, depending on the flow parameters and spacetime properties, the flow may have multiple critical points. The critical point located close to the compact object is termed the inner critical point, while one situated farther away is referred to as the outer critical point. Their respective locations are denoted by $r_{\textrm{in}}$ and $r_{\textrm{out}}$. We therefore proceed to examine the nature and location of the critical points as functions of the flow parameters and the spacetime geometry. Fig.~\ref{fig:energy1} illustrates the variation of the flow energy $(\mathcal{E})$ as a function of the critical
point location $(r_c)$ for different values of $\gamma$. The angular momentum is fixed at $\lambda = 3.4$, and solid (green), dotted (blue) and dashed (magenta) curves represent the
saddle, spiral and nodal type critical points, respectively. The chosen values of $\gamma$ are $1.0$, $0.8$, $0.5$, $0.46$, and $0.4$, as indicated in the Fig.~\ref{fig:energy1}.  We observe that the effect of $\gamma$ is primarily confined to the inner critical points $(r_{\rm in})$, while the outer critical points $(r_{\rm out})$ remain largely unaffected. Moreover, there exists a lower bound on $\gamma$, below which the inner critical point disappears, and only the outer critical point remains in this 
$\gamma$-metric. As in the Schwarzschild case ($\gamma=1.0$), we also find that the outer saddle-type critical points in this spacetime admit an upper bound on their location, beyond which no such critical point exists for the relativistic flow. 

\begin{figure}[h!]
    \begin{center}
    \includegraphics[width=\columnwidth]{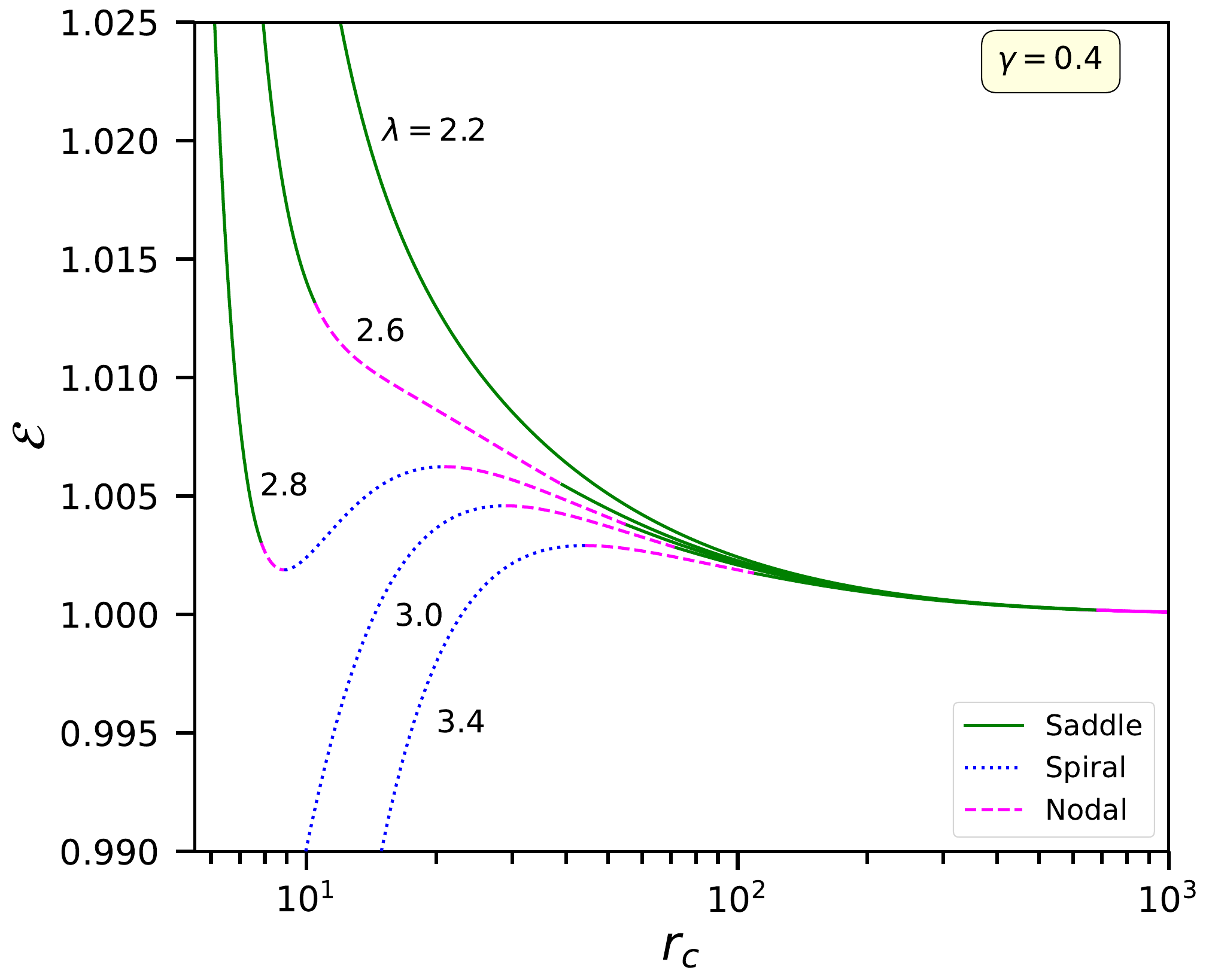}
    \end{center}
    \caption{Variation of energy ($\mathcal{E}$) as a function of critical points $(r_c)$ for $\gamma=0.4$. Here, angular momentum is varied as $\lambda =$ 2.2, 2.6, 2.8, 3.0, and 3.4, respectively. The solid (green), dotted (blue) and dashed (magenta) segments in a given curve denote saddle, spiral, and nodal types critical points, respectively. See the text for further details.
    }
    \label{fig:energy2}
\end{figure}

\begin{figure}[h!]
    \begin{center}
    \includegraphics[width=\columnwidth]{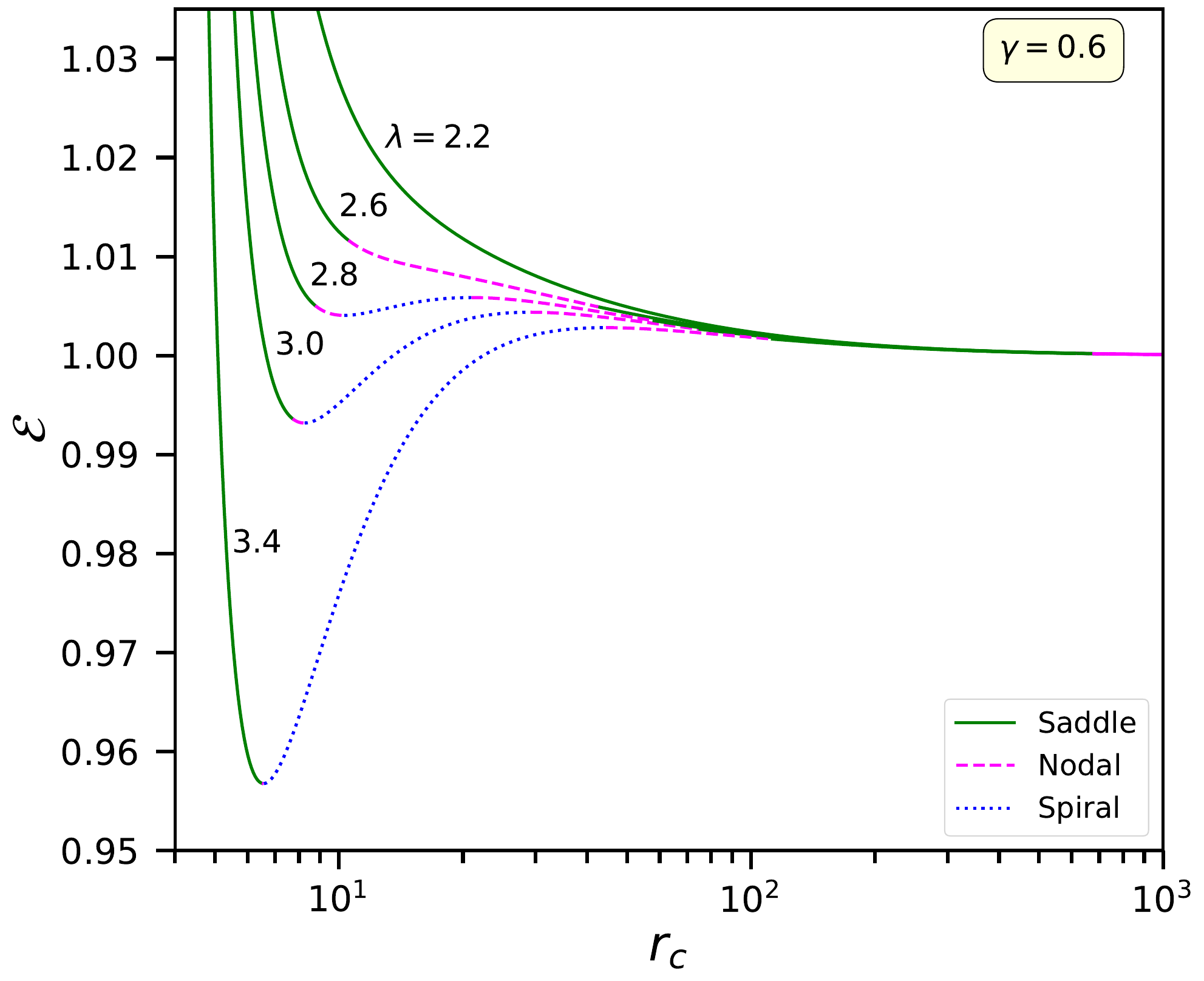}
    \end{center}
    \caption{Variation of the energy ($\mathcal{E}$) with critical point location ($r_c$) for a fixed deformation parameter $\gamma=0.6$. The angular momentum is varied as $\lambda=2.2$, 2.6, 2.8, 3.0, and 3.4, as indicated in the figure. The solid (green), dotted (blue), and dashed (magenta) segments in a given curve denote saddle, spiral, and nodal types critical points, respectively. See the text for details.}
    \label{fig:energy3}
\end{figure}

In Fig.~\ref{fig:energy2} and Fig.~\ref{fig:energy3}, we plot $\mathcal{E}$ as a function of $r_{c}$ for $\gamma=0.4$ and $\gamma=0.6$, respectively. In both plots, the angular momentum of the flow is fixed at $\lambda=2.2, 2.6,2.8, 3.0,$ and $3.4$. For $\lambda = 2.2$ and $2.6$, no multiple critical points are found for either value of $\gamma$ at a given energy. For $\gamma = 0.4$ we get multiple saddle type critical points for $\lambda = 2.8$, which is absent for $\gamma = 0.6$. On the other hand, for $\lambda = 3.0$ and $3.4$, multiple critical points appear for $\gamma = 0.6$, whereas they are absent for $\gamma = 0.4$, where the inner saddle-type branch disappears altogether. This can be attributed to the fact that a decrease in $\gamma$ strengthens the centrifugal barrier experienced by the matter, that prevents to form the inner critical points. This is also evident from the effective potential behaviour shown in Figs.~\ref{fig:fig_poten2} and \ref{fig:fig_poten3}.
 
\begin{figure}[h!]
    \begin{center}
    \includegraphics[width=\columnwidth]{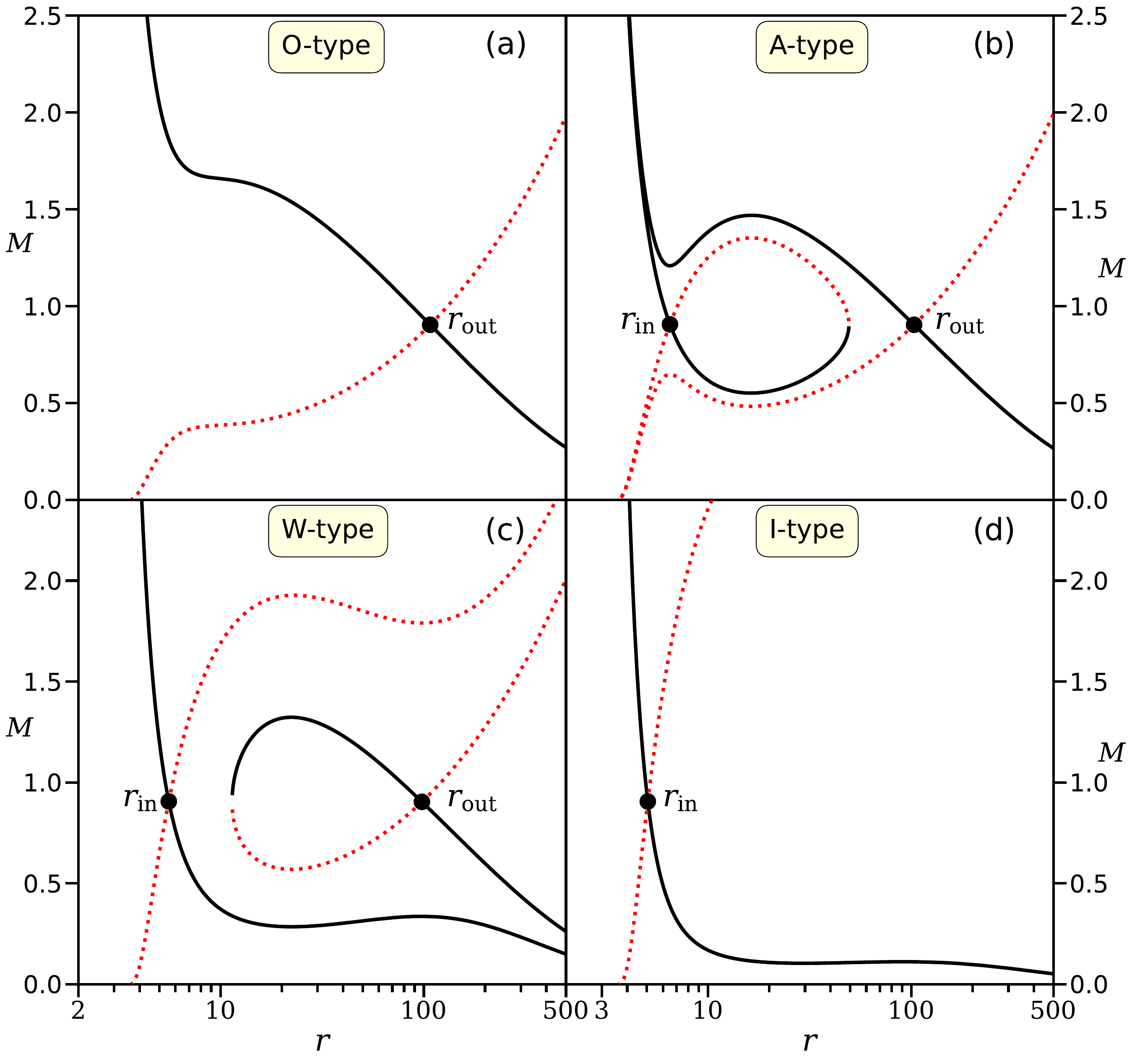}
    \end{center}
    \caption{Different classes of transonic flow solutions ($M-r$ plane) around the $\gamma$-metric for a fixed deformation parameter $\gamma=0.55$. Panels (a)-(d) display the characteristic O-type, A-type, W-type, and I-type solution topologies, respectively. Accretion and wind branches are depicted by solid (black) and dotted (red) curves, while the filled black circles indicate the corresponding critical points ($r_{\rm in}$ and/or $r_{\rm out}$). See the text for the details.}
    \label{fig:solutions}
\end{figure}

After that, we use the flow velocity $v_c$ and temperature $\theta_c$  evaluated at the critical point (with the help of Eq.~\ref{vc} and Eq.~\ref{csc}) as initial conditions and integrate Eqs.~\ref{dvdr} and \ref{dthetadr} from the critical point to $r_{\rm edge}^{\rm out}$ and again from the critical point to $r_{\rm edge}^{\rm in}$. Thereafter, we combine both these branches to
get the global transonic accretion solution. The nature of the accretion solution depends on the position of critical points. When the accretion
flow admits only an inner critical point $r_{\rm in}$, the corresponding flow topology is referred to as I-type, whereas the presence of only an outer critical point $r_{\rm out}$ corresponds to an O-type solution. Moreover, depending on the background geometry and
flow parameters, the accretion flow may contain
both the inner and outer critical points simultaneously. In
such cases, the solution topology is determined by comparing the entropy accretion rate at these critical points. If the entropy accretion rate associated with the inner critical point exceeds that of the outer one, i.e., $\dot{\mathcal{M}}_{\rm in} > \dot{\mathcal{M}}_{\rm out}$, the flow is identified as an A-type solution. Conversely, if $\dot{\mathcal{M}}_{\rm in} < \dot{\mathcal{M}}_{\rm out}$, the corresponding topology is referred to as W-type. These different solution topologies are illustrated in Fig.~\ref{fig:solutions}, where we fix the deformation parameter at $\gamma=0.55$, the specific energy at $\mathcal{E}=1.002$ and vary the angular momentum $\lambda$. In all panels, the black solid and red dotted curves denote the accretion and wind branches, respectively, and the filled black circles mark the critical points. For $\lambda=2.80$, the flow has only an outer critical point at $r_{\rm out}=107.4447$ (Fig.~\ref{fig:solutions}a), which corresponds to an O-type solution. For $\lambda=3.0$, it has an outer critical point at $r_{\rm out}=103.0686$ and an inner one at $r_{\rm in}=6.4936$ (Fig.~\ref{fig:solutions}b), which yield an A-type solution. For $\lambda=3.20$, the critical points lie at $r_{\rm out}=97.8327$ and $r_{\rm in}=5.5664$ (Fig.~\ref{fig:solutions}c), which produces a W-type solution. Finally, for $\lambda=3.40$, the flow has only an inner critical point at $r_{\rm in}=5.0522$ (Fig.~\ref{fig:solutions}d), resulting in an I-type solution.

\section{Parameter space}
\label{sec:parameter_space}
\begin{figure}
    \begin{center}
    \includegraphics[width=\columnwidth]{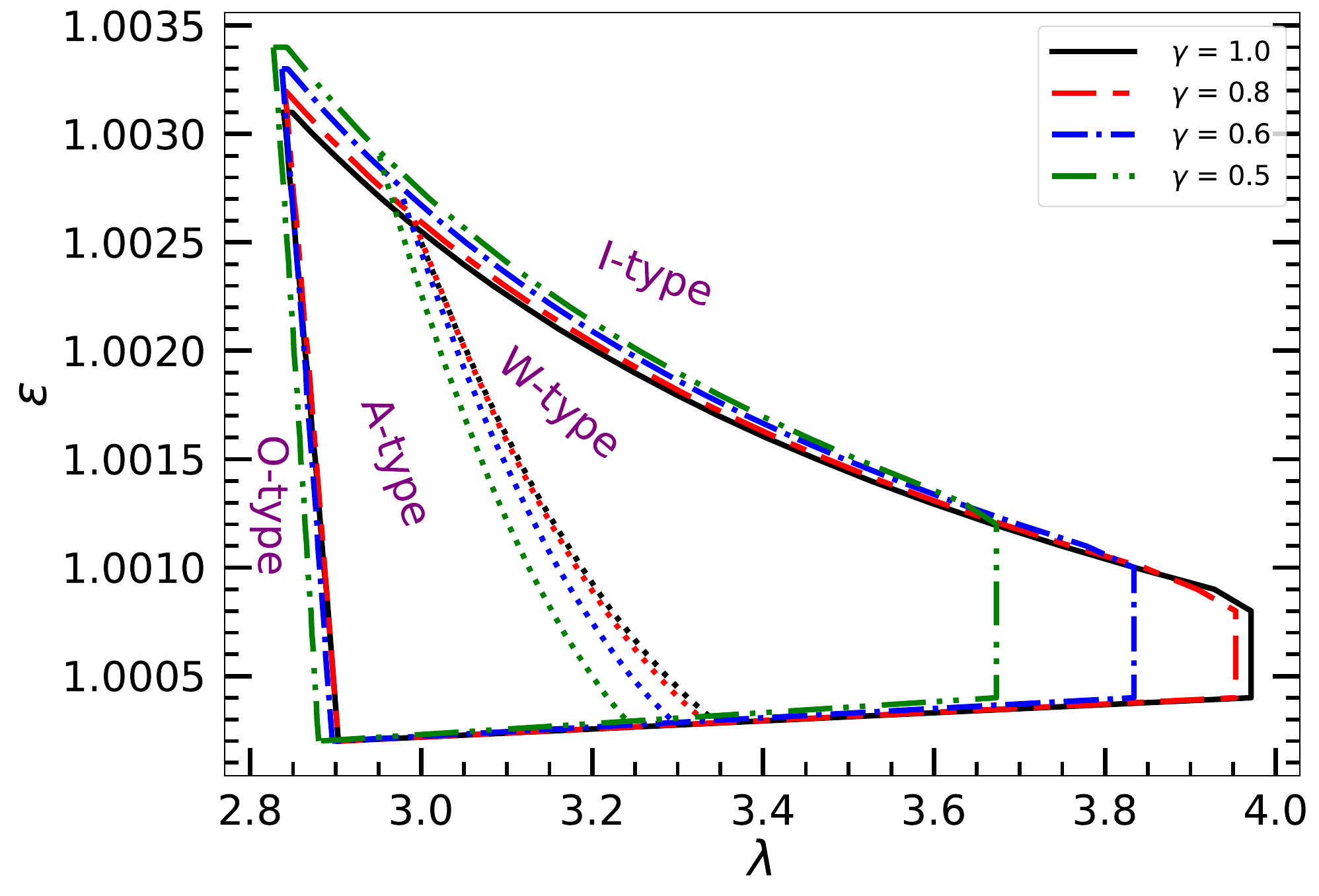}
    \end{center}
    \caption{Parameter space in the $\mathcal{E}$-$\lambda$ plane showing the occurrence of different transonic solution topologies for various values of the deformation parameter $\gamma$. The solid (black), small-big-dashed (red), dot-dashed (blue), and dot-dot-dashed (green) boundaries are obtained for $\gamma=1.0$, 0.8, 0.6, and 0.5, respectively, where multiple saddle-type critical points exist. Based on the global flow topology, four distinct class of solutions, namely O-type, A-type, W-type, and I-type, are identified and marked in the parameter space. See text for details.}
    \label{fig:parameter space}
\end{figure}

In this section, we will show how the parameter space in $\mathcal{E}-\lambda$ plane for the four above mentioned solution types alters depending upon the spacetime deformation  parameter ($\gamma$). Fig.~\ref{fig:parameter space} illustrates the modifications of the parameter space with decreasing 
$\gamma$. The regions enclosed by solid (black), small-big-dashed (red), dot-dashed (blue), and dot-dot-dashed (green) curves correspond to $\gamma=1.0,0.8,0.6$, and $0.5$, respectively.  Each of these regions is further divided by a dotted curve that separates A-type solutions (on the left) from W-type solutions (on the right). The four distinct regions correspond to different solution topologies,
labeled as O-type, A-type, W-type, and I-type in the figure. It is evident from the figure that the effective domain of the parameter space for multiple critical points shrinks as $\gamma$ decreases. 

\section{Radiative properties}
\label{sec:Rad_properties}
In this section, we examine the production of relativistic thermal Bremsstrahlung (free-free) radiation from the accretion flow, as it is one of the primary radiative mechanisms in single temperature accretion flows \cite{Sarkar-Das2016,Okuda-etal2019}. This emission mechanism arises when electrons undergo acceleration in the Coulomb field of heavy ions and leads to photon emission. In case of the low angular momentum flow, the flow temperature can reach upto $\sim 10^{12}$ K close to the compact objects, which makes flow relativistic ($k_B T_e \gtrsim m_e c^2$). Therefore, in this regime, we cannot neglect the contribution of the electron-electron interactions in addition to the electron-ion interactions. To account these effects, we adopt a relativistically corrected Bremsstrahlung emissivity ($\varepsilon_\nu$) following \cite{Novikov-Thorne1973}. Therefore, the expression for this emissivity at frequency $\nu$ is given by,
\begin{align}\label{eq:emissivity}
\varepsilon_\nu = & \frac{32 \pi e^6}{3 m_e c^3}
\left(\frac{2 \pi}{3 m_e k_B}\right)^{1/2}
n_e n_i Z_i^2 T_e^{-1/2} \nonumber\\
 &\times \left(1 + 4.4\times 10^{-10} T_e \right)
\exp\left(-\frac{h\nu}{k_B T_e}\right)\bar{g}_b ,
\end{align}
where $h$ and $k_B$ denote the Planck and Boltzmann constants, respectively. The quantities $e$, $m_e$, $n_e$, and $T_e$ represent the electron charge, mass, number density, and temperature, while $n_i$ and $Z_i$ correspond to the ion number density and ion's charge. The factor $\bar{g}_b$ is the velocity-averaged Gaunt factor that takes into account the quantum corrections. Here, it is assumed to be constant with a value of 1.2~\cite{Yarza:2020loo}. Because of the charge neutrality condition, the electron and ion number densities are identical ($n_e = n_i$) in the flow. Owing to the much smaller mass of electrons compared to ions, electrons are less efficiently heated and results to a lower temperature than that of ions. Therefore, the overall flow temperature $T$ closely follows the ion temperature ($T_i$). Throughout this work, we assume a single-temperature accretion flow. Accordingly, the electron temperature is approximated following \cite{Chattopadhyay-Chakrabarti2002} as $T_e=\sqrt{m_e/m_i}T$, where $m_i$ is the ion mass.
The intense gravitational field of the central compact object causes the emitted photons to undergo gravitational redshift before reaching a distant observer. In addition, the rotational motion of the accretion flow introduces a Doppler shift in the observed radiation. The emitted ($\nu_e$) and observed frequencies ($\nu_o$) are related as follows~\cite{Luminet1979}:
\begin{equation}\label{eq:redshift}
        \nu_e=\nu_o u^t \left(1+\frac{r \Omega}{c} \sin \theta_0\sin\phi\right).
    \end{equation}

\begin{figure}[h!]
    \begin{center}
    \includegraphics[width=\columnwidth]{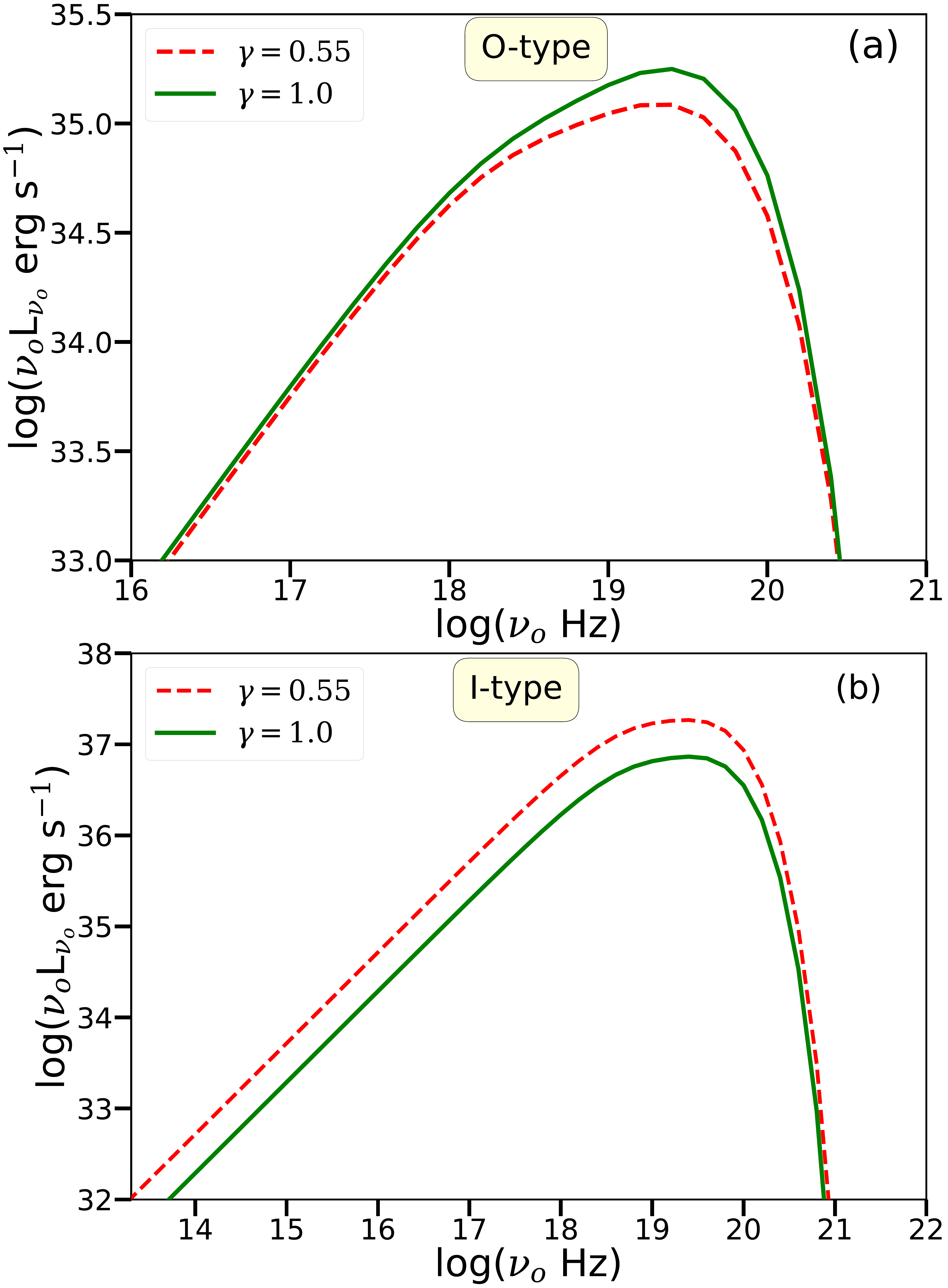}
    \end{center}
    \caption{Spectral energy distribution for (a) O-type solutions and (b) I-type solutions. For O-type solutions, we consider ($\mathcal{E},\lambda=1.0005,2.8$) and for I-type solutions, we consider ($\mathcal{E},\lambda=1.003,3.4$). In both panels, the dashed (red) and solid (green) curves denote results obtained for $\gamma= 0.55$ and $1.0$, respectively. See the text for details.
    }
    \label{fig:O_I-spectra}
\end{figure}

Here, $\theta_0$ is the inclination angle of the accretion disk with respect to the line of sight of an observer at infinity, and here we consider $\theta_0=45^\circ$. Therefore, we can write the monochromatic disk luminosity $(L_{\nu_o}$ in $~{\rm{erg}~ \rm{s}^{-1} \rm{Hz}^{-1} }$) for an asymptotic observer as, 

\begin{align}\label{eq:spec}
L_{\nu_o}=&2\int_{r^{\rm in }_{\rm edge}}^{r^{\rm out}_{\rm edge}}\int_{0}^{2\pi} \varepsilon_{\nu_e} H r~dr d\phi ,\nonumber\\
=&4.875 \times10^{10} \bar{g_b}\int_{r^{\rm in }_{\rm edge}}^{r^{\rm out}_{\rm edge}}\int_{0}^{2\pi} (1+4.4 \times 10^{-10}T_e)  \rho^2 T_e^{-1/2}\nonumber\\
&\times {\rm exp} \left(\frac{-h\nu_o u^t \left(1+\frac{r \Omega}{\sqrt{2}c}\sin\phi\right)}{k_B T_e}\right) H r ~dr d\phi,
\end{align}
where $r^{\rm in}_{\rm edge}$ and $r^{\rm out}_{\rm edge}$ are the inner and $r_{\rm edge}^{\rm out}$ of the accretion disk, respectively. In our analysis, $r^{\rm out}_{\rm edge}$ is defined as the radius at which $H/r^{\rm out}_{\rm edge} \sim 1$. From previous studies, we know that among all four types of solutions (O, A, W, and I), the I-type solutions can produce higher luminosities~\cite{Patra-etal2023, Sen-etal2024}.  In our analysis, we consider ADM mass of the spacetime to be $M_\bullet=10 M_\odot$ and mass accretion rate of $\dot{M}=0.1 \dot{M}_{\rm EDD}$. 
In Fig.~\ref{fig:O_I-spectra}a, we illustrate the variation of spectral energy distribution (SED) for two different $\gamma$ parameters for O-type solutions. The dashed (red) and solid (green) curves represent the SEDs corresponding to $\gamma=0.55$ and 1.0, respectively. Here, we choose the energy parameter $\mathcal{E} = 1.0005$ and angular momentum $\lambda = 2.8$. For both cases, the accretion solutions pass through the outer critical points $i.e.,$ O-type solutions. Fig.~\ref{fig:O_I-spectra}b represents the variation of SED for $\gamma=0.55$ (red dashed curve) and $\gamma=1.0$ (green solid curve) for I-type solutions. Here, we choose the energy parameter $\mathcal{E} = 1.003$ and angular momentum $\lambda = 3.4$. The solutions only possess the inner critical points. As we decrease the $\gamma$ parameter, the critical points ($r_{\rm c}$) (inner or outer) shift in the outward direction, but the effective distance of critical point $\sqrt{g_{\phi \phi}}|_{r_{\rm c}}$ from the central singularity at $\pi/2$ plane  is greater for the outer critical point for $\gamma=0.55$ than that of $\gamma=1.0$ (see Table~\ref{tab:table1}). On the other hand, the effective distance of critical point $\sqrt{g_{\phi \phi}}|_{r_{\rm c}}$ from the central singularity is smaller for the inner critical point for $\gamma=0.55$ than that of $\gamma=1.0$. Therefore, solutions with a smaller effective critical radius, corresponding to a critical point located closer to the central compact object and produce higher luminosity.

\begin{table}
\caption{Critical point locations and their natures obtained for various sets of input parameters.}

\begin{center} 
\label{tab:table1}
		
\begin{tabular}{@{\hspace{0.1cm}} c @{\hspace{0.2cm}} c @{\hspace{0.2cm}} c @{\hspace{0.2cm}} c @{\hspace{0.2cm}} l @{\hspace{0.2cm}} c @{\hspace{0.2cm}} c
@{\hspace{0.2cm}} c
@{\hspace{0.1cm}} }\hline\hline
			
\multicolumn{3}{r}{Input Parameters} & &  \multicolumn{3}{r}{Critical Points} &\\	

			\cline{1-3}
			\cline{5-8}

$\lambda$ & $\mathcal{E}$ & $\gamma$  &&   & \multicolumn{1}{l}{Location} & Effective distance & Type \\ 
&  &  & &   & \multicolumn{1}{c}{($r_{\rm c}$)} & ($\sqrt{g_{\phi \phi}}|_{r_{\rm c}, \theta=\pi/2}$) & \\ \hline

     &           &      && $r_{\rm in}$  & $4.7865$ & $4.7865$ & Saddle \\
3.4 & 1.003 & 1.0  && $ $ &  &    &  \\
     &        &      && $r_{\rm out}$ & -- &    --  & -- \\ \hline

&           &      && $r_{\rm in}$  & $5.0447$ & $3.78579$ & Saddle \\

3.4 & 1.003 & 0.55  && $ $ &  &    &  \\
     &        &      && $r_{\rm out}$ &--  & --  & -- \\ \hline
&           &      && $r_{\rm in}$  & -- & -- & -- \\
     2.8 & 1.0005 & 1.0  && $ $ &  &   &  \\
     &        &      && $r_{\rm out}$ & 343.4631     &343.4631 &  Saddle\\ \hline
&           &      && $r_{\rm in}$  & -- &--& -- \\
     2.8 & 1.0005 & 0.55  && $ $ &  &    &  \\
     &        &      && $r_{\rm out}$ & 345.1175      &344.2960  & Saddle\\ \hline

		\end{tabular}
	\end{center}
\end{table}

\begin{figure}[h!]
    \begin{center}
    \includegraphics[width=\columnwidth]{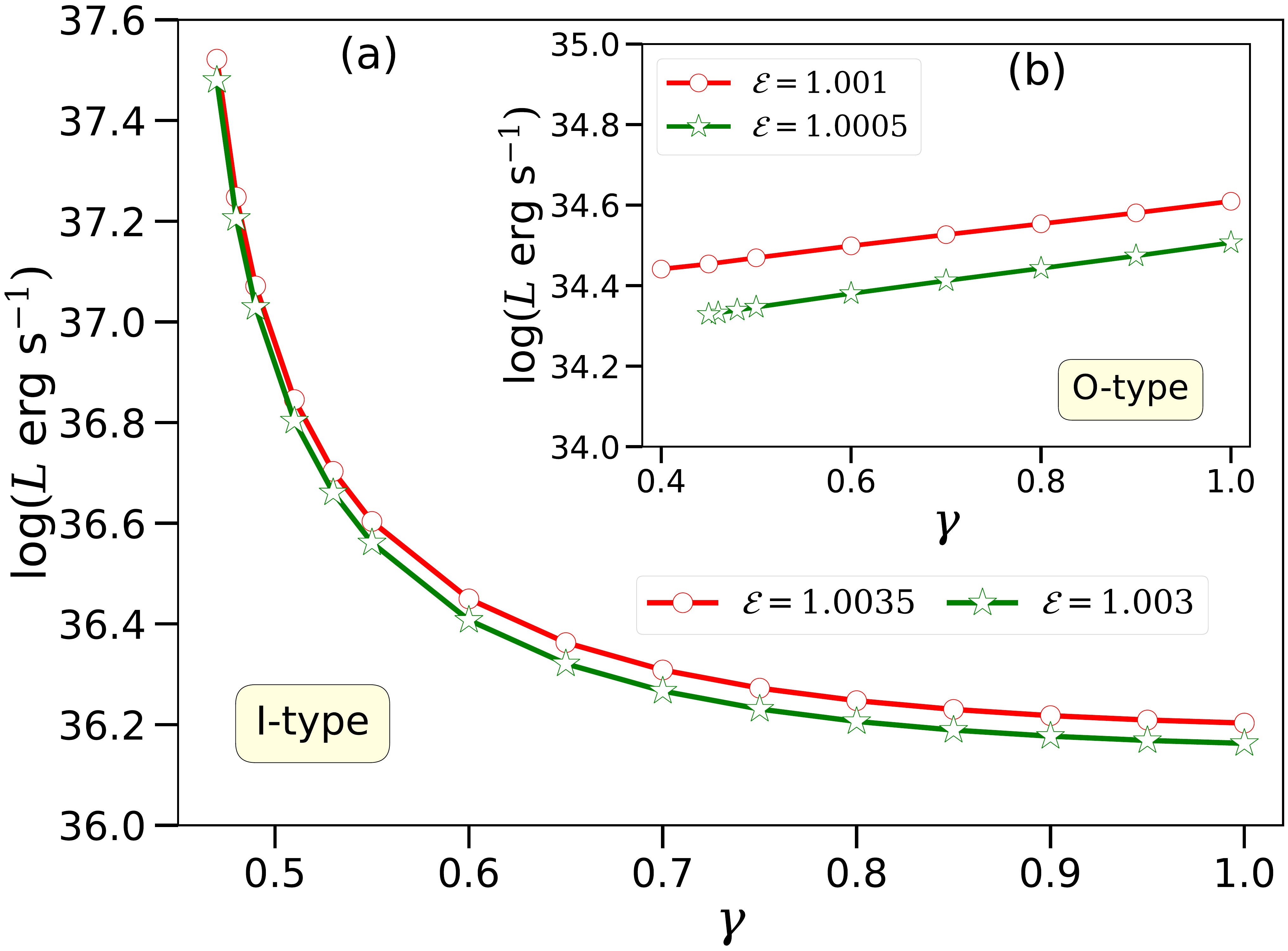}
    \end{center}
    \caption{Bolometric luminosity ($L$) of the accretion flow as a function of the deformation parameter $\gamma$. Panel (a) (main plot) shows the I-type solutions with $\lambda = 3.4$, and panel (b) (inset) shows the O-type solutions with $\lambda = 2.8$. Circles (red) and asterisks (green) joined with lines denote results corresponding to $\mathcal{E} = 1.0035$ and $1.003$ in panel (a), and to $\mathcal{E} = 1.001$ and $1.0005$ in panel (b), respectively. See the text for details.}
    \label{fig:O_I-lumi}
\end{figure}

Now, we want to explore the variation of bolometric disk luminosity with $\gamma$ parameters. For this reason, we integrate the monochromatic luminosity (Eq.~\ref{eq:spec}) over all frequency ranges, as follows:
\begin{equation}
    \label{eq:bolo-lumin}
    L=\int_0^\infty L_{\nu_o} d\nu_o.
\end{equation}
Fig.~\ref{fig:O_I-lumi}a (main plot) shows the variation of bolometric disk luminosity with $\gamma$ parameters for I-type solutions. Here, we consider $\lambda=3.4$, and the red circle and green asterisks joining lines denote $\mathcal{E}=1.0035$ and $1.003$, respectively. In Fig.~\ref{fig:O_I-lumi}b (inset plot) we have plotted the variation of luminosity with $\gamma$ parameters for O-type solutions with $\lambda=2.8$. The red circle and green asterisks joining lines represent for $\mathcal{E}=1.001$ and $1.0005$, respectively. For O-type solutions, we get that the luminosity increases as we increase $\gamma$ parameters. Whereas, for I-type solutions, we see that the luminosity decreases as we increase $\gamma$ parameters.  This also happens as the effective distance of the outer critical points is smaller for higher $\gamma$, as is opposite for the I-type solutions (see Table~\ref{tab:table1}).

Now, we want to see the image of the hot accretion disk in the $\gamma$-metric. The electromagnetic flux radiated from the advective accretion flow at a radial distance $r$ is given by,

\begin{align}\label{eq:flux}
F_{\rm emi}(r)=&2 \int_0^\infty\varepsilon_{\nu_e} H d\nu_e,\nonumber\\
=&4.875 \times10^{10} \bar{g_b}  \rho^2 T_e^{-1/2}\times (1+4.4 \times 10^{-10}T_e)H \nonumber\\
&\times\int_0^\infty {\rm exp} \left(\frac{-h \nu_e}{k_B T_e}\right) d\nu_e.
\end{align}
 As photons emitted from the accretion disk propagate through the curved spacetime, they experience gravitational lensing, which allows only a fraction of light rays to reach a distant observer. Therefore, to construct the corresponding intensity map on the observer's sky, the photon trajectories are traced backward in time by integrating the null geodesic equations from the observer's position $(r_0, \theta_0)$ to their emission points $(r_{\rm emi}, \pi/2)$ on the accretion disk.
We solve the null geodesic equations and calculate the corresponding redshift factor at that location where the photon hits the accretion disk.
The static and axisymmetry of the spacetime  described by the metric in Eq.~\ref{eq:line} imply the existence of two conserved quantities along the photon trajectories, namely the energy 
$E_{\rm ph}$
and the angular momentum 
$L_{\rm ph}$. Consequently, the 
$t$, $\phi$, $r$ and $\theta$ components of the null geodesic equations can be expressed as,
\begin{align}
\label{Eq:geodesics}
\dot{t}=&-\frac{E_{\rm ph}}{g_{tt}},~~
\dot{\phi}=\frac{L_{\rm ph}}{g_{\phi\phi}},\nonumber\\
\ddot{r}=&-\Gamma^r_{\mu\nu}\dot{x}^{\mu}\dot{x}^{\nu},~~
\ddot{\theta}=-\Gamma^\theta_{\mu\nu}\dot{x}^{\mu}\dot{x}^{\nu},\end{align}
where, $\Gamma^r_{\mu\nu}$ and $\Gamma^\theta_{\mu\nu}$, are the Christoffel symbols of this spacetime.  

The observed flux for a distant observer is given by,
\begin{equation}
F_{\rm obs}(r)=g^4 F_{\rm emi}(r).
\end{equation}
Here, $g$ is the redshift factor, which is given by,
\begin{equation}
g=\frac{k_{\mu}u^{\mu}_0}{k_{\mu}u^{\mu}_e}=\frac{1}{u^t(1-b\Omega)-\frac{v\gamma_v\sqrt{\frac{B(r,\pi/2)}{A(r)}}k^r}{E_{\rm ph}}}.
\end{equation}

Here $b=L_{\rm ph}/E_{\rm ph}$ and $u^\mu_0$, $u^\mu_{\rm e}$ and $k^\mu$ are respectively, the four-velocity of the observer at infinity, the flow velocity of the accreting matter at the emitting point on the accretion disk and the four-velocity of the photon. The ray-tracing procedure begins by specifying the observer's position $(r_0, \theta_0)$ and the image-plane coordinates $(x_0,y_0)$. These quantities uniquely determine the corresponding $b$ or equivalently, the constants of motion $L_{\rm ph}$ and $E_{\rm ph}$.

Here, we consider that the inner edge ($r_{\rm edge}^{\rm in}$) of the accretion disk is at ($r_{\rm s}+0.0001$) and $r_{\rm edge}^{\rm out}$ is located at $20 r_{\rm g}$. 
Now, if we fix the observer's location $i.e.,$ $r_0$ and $\theta_0$, we can find out the $L_{\rm ph}, E_{\rm ph}, \dot{r},$ and $\dot{\theta}$ of photon at the observer's location as follows~\cite{Paul-etal2020}: 
\begin{align}
\label{Eq:initial}L_{\rm ph}=&\frac{-\sqrt{g_{\phi \phi}}x_0}{r_0}\Bigg|_{(r_0,\theta_0)},~~
E_{\rm ph}=\sqrt{-g_{\rm tt}}\left(1-\frac{L_{\rm ph}}{g_{\phi\phi}\sqrt{-g_{\rm tt}}}\right)\Bigg|_{(r_0,\theta_0)},\nonumber\\
\dot{r}=&-\frac{1}{r_0\sqrt{g_{\rm rr}}}\sqrt{r_0^2-x_0^2-y_0^2}\Bigg|_{(r_0,\theta_0)},~~
\dot{\theta}=\frac{y_0}{r_0\sqrt{g_{\theta\theta}}}\Bigg|_{(r_0,\theta_0)}.\end{align}

To trace back all the photon trajectories, we have to solve two first order and two second order differential equations. Therefore, we need six initial conditions, along with the initial values of $E_{\rm ph}$ and $L_{\rm ph}$, which are constant throughout the photon's geodesic. In our whole analysis we consider $r_0=10^3~r_{\rm g}$, $t_0=0$, and $\phi_0=0$ and different $\theta_0$. Along with these four initial conditions we can calculate $\dot{r}|_{(r_0,\theta_0)}$ and $\dot{\theta}|_{(r_0,\theta_0)}$ using Eq.~\ref{Eq:initial} for a particular apparent observed direction ($x_0, y_0$) on the image plane. Once all the required initial conditions are specified, the coupled differential equations are numerically integrated backward from the observer toward the lensing object. For each selected point $(x_0,y_0)$ on the observer's image plane, this procedure determines a unique photon geodesic. By varying $(x_0,y_0)$ while keeping the observer's spacetime coordinates $(t_0,r_0,\theta_0,\phi_0)$ fixed, different photon trajectories can be generated. For each point $(x_0,y_0)$ on the observer's image plane, the corresponding photon geodesic is back traced toward the lensing object. If the trajectory intersects the accretion disk at a radial position $r_{\rm emi}$, the observed redshifted flux $F_{\rm obs}(r_{\rm emi})$ is assigned corresponding to ($x_0,y_0$). In contrast, if the photon fails to reach the disk, zero flux is assigned, resulting in a dark region in the image. By repeating this procedure over the entire image plane, the intensity map of the accretion disk is constructed for a fixed observer position.

\begin{figure*}[t]
    \begin{center}
    \includegraphics[width=2\columnwidth, height=\textwidth,keepaspectratio]{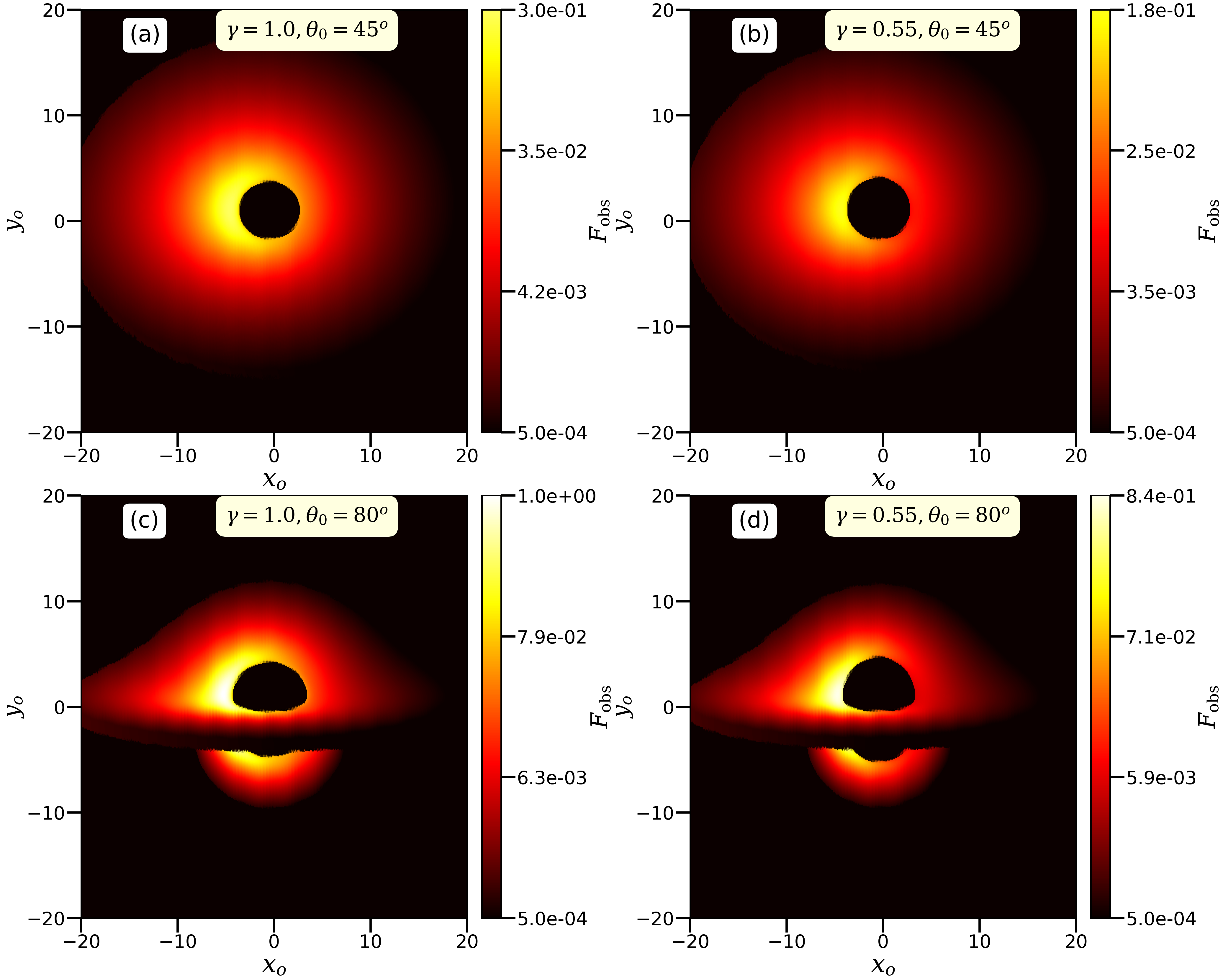}
    \end{center}
    \caption{Images of hot accretion disks for $\gamma=1.0$ and $\gamma=0.55$, obtained for $\mathcal{E}=1.0005$ and $\lambda=2.8$ (O-type solutions). The disk extends up to $r=20\,r_{g}$, and results are shown for observer inclination angles $\theta_{0}=45^\circ$ and $80^\circ$. The image intensity is normalized by the maximum observed flux in the $\gamma=1.0$, $\theta_{0}=80^\circ$ configuration to enable a consistent comparison. See the text for details.
    }
    \label{fig:O_image1}
\end{figure*}

Fig.~\ref{fig:O_image1} presents the images of the accretion disk around the $\gamma$-metric spacetime for different values of the deformation parameter $\gamma$ and observer inclination angle $\theta_0$ for O-type solutions. Panels (a), (b), (c), and (d) correspond to $(\gamma=1.0,\theta_0=45^\circ)$, $(\gamma=0.55,\theta_0=45^\circ)$, $(\gamma=1.0,\theta_0=80^\circ)$, and $(\gamma=0.55,\theta_0=80^\circ)$, respectively. For all panels, we consider energy ($\mathcal{E}$)= $1.0005$ and angular momentum ($\lambda$)= $2.8$. The colorbars indicate the observed flux distribution across the image plane. For a consistent comparative analysis, the observed flux in all four panels is normalized with respect to the maximum flux obtained for the case $(\gamma=1.0,\theta_0=80^\circ)$. By comparing the four panels, we find that, for $\theta_0=80^\circ$, the observed radiation flux from regions close to the central compact object is higher than that obtained for $\theta_0=45^\circ$. Furthermore, the observed flux is higher for $\gamma=1.0$ than for $\gamma=0.55$. This behavior is consistent with our earlier results for the luminosity and spectral energy distribution, both of which are lower for $\gamma=0.55$ compared to the $\gamma=1.0$ case. Thus, the accretion disk images independently support the same trend observed in the luminosity and spectral properties. 

\begin{figure*}[t]
    \begin{center}
    \includegraphics[width=2\columnwidth, height=\textwidth,keepaspectratio]{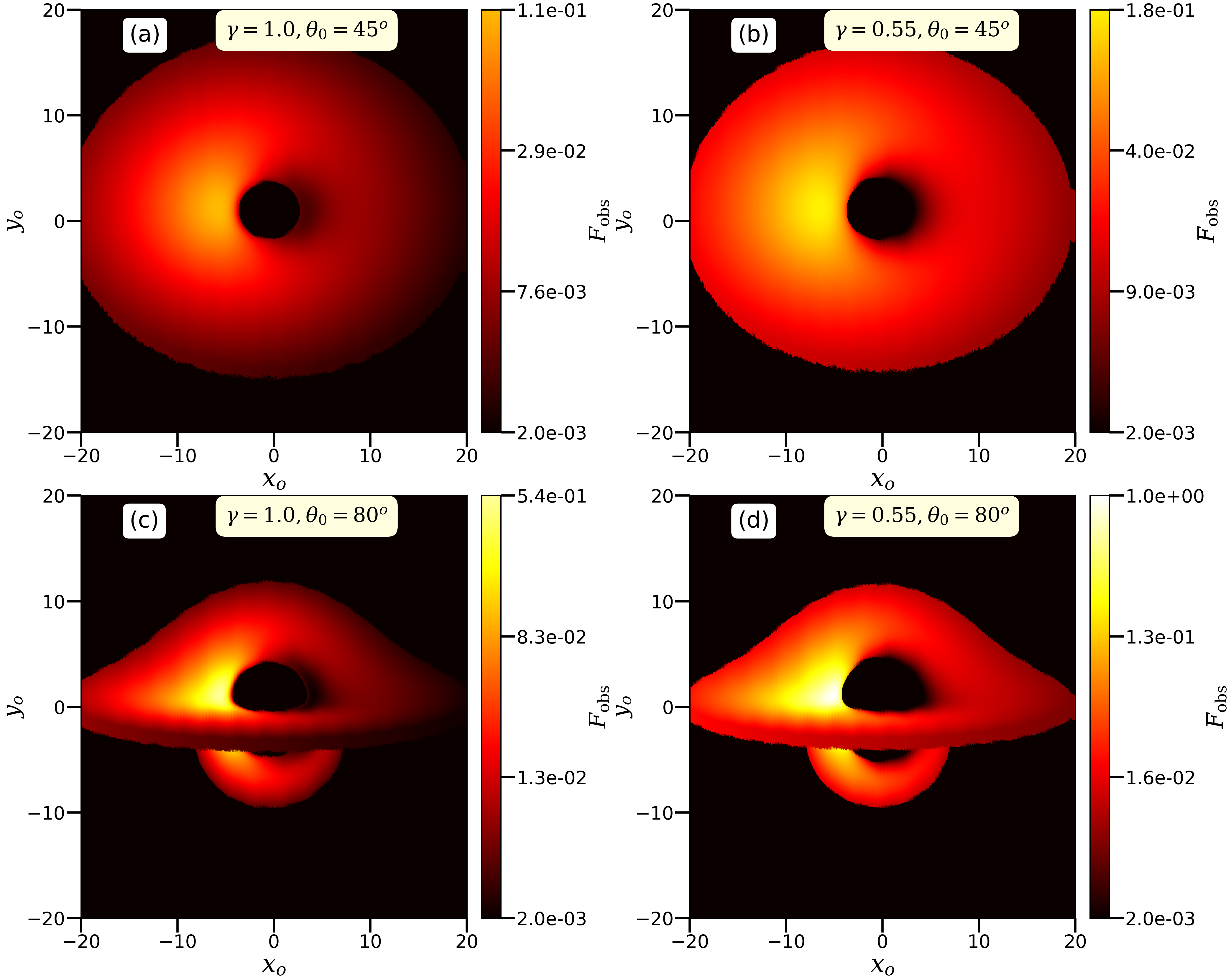}
    \end{center}
    \caption{Synthetic images of hot accretion disks around $\gamma$ metric compact objects for $\gamma=1.0$ and $\gamma=0.55$. The adopted flow parameters, $\mathcal{E}=1.003$ and $\lambda=3.4$, correspond to the I-type solution topology. The disk extends up to an outer radius of $r=20\,r_g$, and images are presented for observer inclination angles $\theta_0=45^\circ$ and $80^\circ$. The intensity scale is normalized by the maximum observed flux in the $\gamma=0.55$, $\theta_0=80^\circ$ configuration, enabling a consistent comparison. See the text for details.
    }
    \label{fig:I-image1}
\end{figure*}

In Fig.~\ref{fig:I-image1}, we portray the accretion disk image around the $\gamma$-metric spacetime for different values of the deformation parameter $\gamma$ and observer inclination angle $\theta_0$ for I-type solutions. Panels (a), (b), (c), and (d) correspond to $(\gamma=1.0,\theta_0=45^\circ)$, $(\gamma=0.55,\theta_0=45^\circ)$, $(\gamma=1.0,\theta_0=80^\circ)$, and $(\gamma=0.55,\theta_0=80^\circ)$, respectively. Here, we consider $\mathcal{E}=1.003$ and $\lambda=3.4$. The colorbars indicate the normalized observed flux distribution on the image plane. We have normalized with respect to the maximum flux obtained for the case $(\gamma=0.55,\theta_0=80^\circ)$. Among the four panels, for ($\gamma=0.55, \theta_0=80^\circ$), the observed radiation flux reaches highest values. Therefore, this suggests that we get luminosity for higher inclination angle and lower $\gamma$ value for I-type solutions. This trend is also consistent with our earlier findings, where both the luminosity and the spectral energy distribution were found to be lower for 
$\gamma=1.0$ than for the 
$\gamma=0.55$ case.

\begin{figure}[h]
    \begin{center}
    \includegraphics[width=\columnwidth]{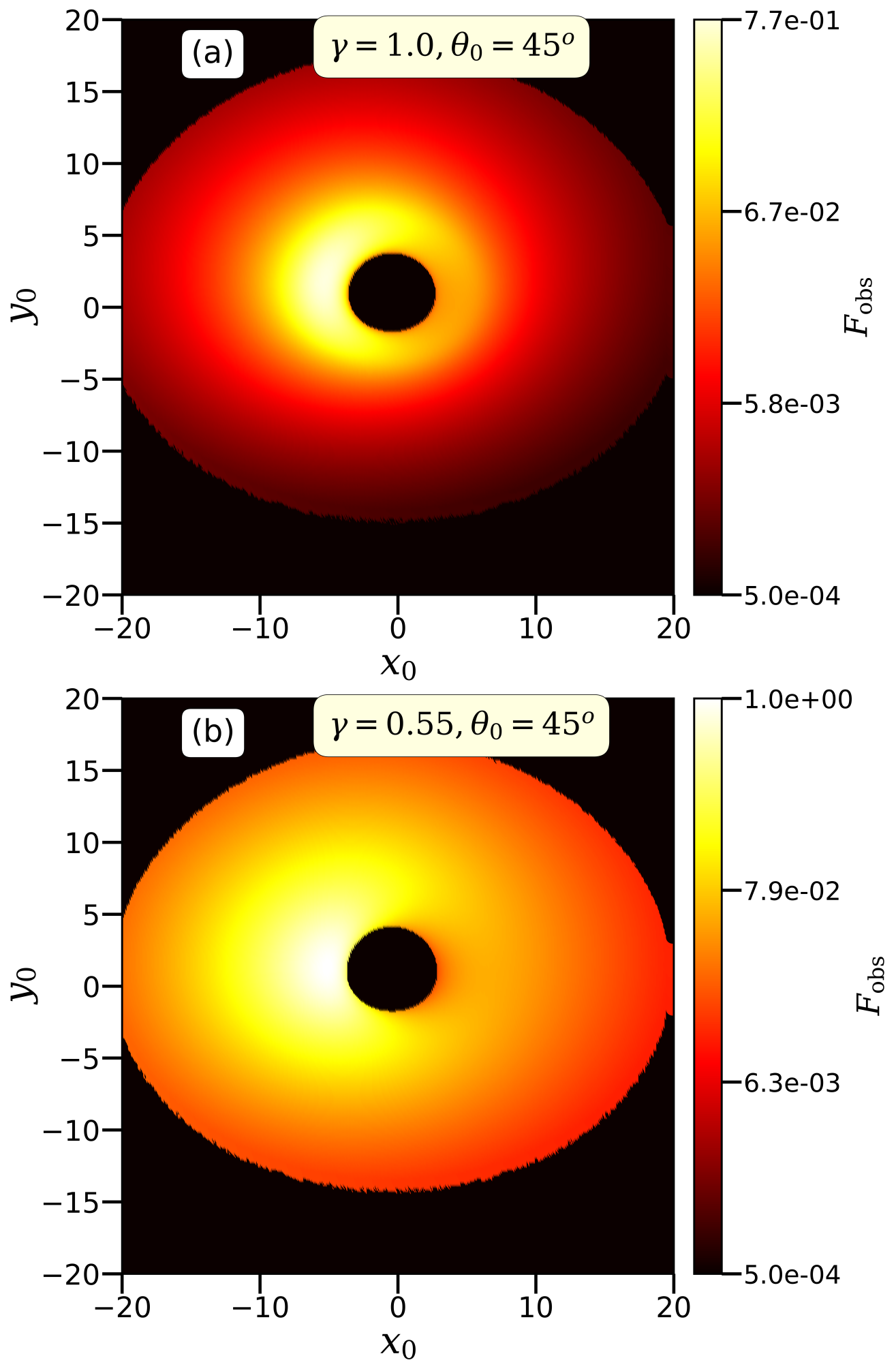}
    \end{center}
    \caption{Comparison of synthetic accretion disk images in two spacetimes characterized by $\gamma=1.0$ and $\gamma=0.55$. The flow parameters are fixed at $\mathcal{E}=1.0013$ and $\lambda=3.11$, corresponding to an A-type solution for $\gamma=1.0$ and a W-type solution for $\gamma=0.55$. The outer edge of the disk is located at $r=20\,r_g$, and both images are computed for an observer inclination angle of $\theta_0=45^\circ$. The displayed flux is normalized to the maximum intensity obtained for the $\gamma=0.55$ model. See the text for details.
    }
    \label{fig:A-W-image}
\end{figure}

Finally, in Fig.~\ref{fig:A-W-image}, we plot the accretion disk image for $\gamma=1.0$ (a) and $\gamma=0.55$ (b) for same flow properties $(\mathcal{E}=1.0013,\lambda=3.11)$ and inclination angle $(\theta_0=45^{\circ})$. The colorbars are normalized with respect to the maximum flux for the case $\gamma=0.55$. For these set of parameters, we get A-type solutions for $\gamma=1.0$ and W-type solutions for $\gamma=0.55$ (see Fig.~\ref{fig:parameter space}) . As, for A-type (W-type) solution, the accretion flow passes through the outer (inner) critical point (see Fig.~\ref{fig:solutions}), the flow density and temperature are higher for W-type solution than A-type solution. As a result, the observed flux from the accretion disk is higher for $\gamma=0.55$ than that of $\gamma=1.0$.

\section{Summary and Conclusions}
\label{sec:Conclusions}

In this work, we have investigated the properties of a steady, inviscid, axisymmetric, advective accretion flow around the Zipoy-Voorhees ($\gamma$-metric) spacetime, one of the simplest static and axisymmetric generalizations of the Schwarzschild spacetime. By solving the relativistic hydrodynamic conservation equations for accreted matter  around this background, we have obtained transonic accretion solutions and examined how the $\gamma$ parameter, which measures deformation from the Schwarzschild black hole, influences the flow dynamics, resulting radiative signatures, and the synthetic images. Our key findings are summarized below.

We show the dependencies of effective potential on $\gamma$ parameter and angular momentum of the flow $\lambda$. As $\gamma$ decreases from $1$ (the Schwarzschild one), the centrifugal barrier in the inner region of the flow becomes stronger and the effective potential deviates from that of the Schwarzschild spacetime. The critical point analysis shows that the $\gamma$ parameter affects the inner critical point ($r_{\rm in}$), while the outer critical point ($r_{\rm out}$) remains mostly unaffected. We also find that below a certain value of $\gamma$ $r_{\rm in}$ ceases to exist, whereas the $r_{\rm out}$ exists. Therefore, $\gamma$ parameter also governs the accretion flow topologies. As $\gamma$ decreases, the strengthened centrifugal barrier suppresses the formation of multiple critical points, therefore, the parameter space for A-type and W-type solutions shrinks. 

From the radiative properties (spectra and luminosity) of the flow, it is evident that the $\gamma$ parameter affects the bolometric luminosity in opposite ways depending on the flow topology. For O-type solutions, the spectra and luminosity increases with increasing $\gamma$, whereas for I-type solutions these decrease with increasing $\gamma$. These behaviour reflects the facts of the effective distance of the relevant critical point from the central compact object changes with $\gamma$. The solutions whose critical points locate close to the compact object produce higher luminosities. Since this effective distance decreases with $\gamma$ for the outer critical point but increases with $\gamma$ for the inner critical point, the two topologies show opposite luminosity trends. This pattern is consistent across both the spectral energy distributions and the bolometric luminosities.

Finally, using general relativistic ray tracing combined with a relativistically corrected thermal bremsstrahlung emissivity, synthetic images of the hot accretion flow as seen by a distant observer have been constructed. For O-type solutions, the observed flux is higher for $\gamma = 1.0$ (Schwarzschild) than for $\gamma = 0.55$, which is consistent with the luminosity and spectral results. On the other hand, for I-type solutions, the observed flux is higher for $\gamma = 0.55$ than for $\gamma = 1.0$ (Schwarzschild). After that, we compare the accretion disk images for identical energy and angular momentum but different flow topology, an A-type solution for $\gamma = 1.0$ versus a W-type solution for $\gamma = 0.55$. 
We find that the W-type solution produces a higher observed flux, since gas accreting through the inner critical point has higher density and temperature than gas accreting through the outer critical point. Together, these results show that the accretion disk images independently confirm the trends found in the luminosity and spectral analyses, and that both $\gamma$ and the accretion flow topology leave clear, potentially observable imprints on the horizon-scale image. Although, the horizonless compact object cast shadows similar to black holes for $0.5\leq \gamma< 1$, which has been shown to be consistent with the observed shadow radius of both Sgr A$^*$ and M87$^*$, the accretion flow topology and the related observational features could differ from that of the Schwarzschild black hole.

However, it should be noted that our present analysis is based on several simplifying assumptions such that we considered a single temperature, non-magnetized accretion flow in which thermal bremsstrahlung is the sole radiative mechanism. In reality, synchrotron and inverse Compton emission usually dominate, even in presence of a weak magnetic field, and a two-temperature treatment becomes important close to the compact object. Moreover, we haven't considered any kind of dissipative term to capture angular momentum transport, or the time variability of the flow properties. Even with these simplifications, this method shows how the deformation parameter $\gamma$ affects the accretion flow and its image, without the extra complexity of magnetic fields, viscous terms, or high computational costs. Based on this, we plan to extend the present framework in future work by incorporating magnetic fields, non-thermal emission, and a two-temperature plasma. 

As a first step in this direction, a polarization study is already in progress, in which we compute the polarized synchrotron emission. In this work, we aim to examine how the deformation parameter modifies the degree of linear polarization fraction, the electric vector position angle pattern across the image, and the circular polarization arising from Faraday rotation. In particular, we want to test whether the $\gamma$-metric shows a polarimetric signature that is distinct from the Schwarzschild case, even when the total intensity images look similar. Such a study would also help to separate the effect of the spacetime geometry from that of the magnetic field configuration. Finally, a comparison with the polarimetric observations of M87$^*$ and Sgr A$^*$ by the EHT might allow us to better constrain the admissible range of $\gamma$.

\section*{Acknowledgments}

GS gratefully acknowledges the warm hospitality of the PAMU, ISI Kolkata, during her visit, where most of this work was carried out. She also extends her gratitude to the IUCAA Pune, for the support and the vibrant academic atmosphere provided while part of this work was done. The work of RS is supported by a Start-Up research grant provided by ISI Kolkata, India.

\newpage
\bibliography{thesisref}
\end{document}